\documentclass[aps,prd,reprint,amsmath,amssymb,nofootinbib,floatfix]{revtex4-2}
\usepackage{orcidlink}
\usepackage[ruled,vlined]{algorithm2e}
\usepackage{cancel}
\usepackage{cleveref}

\usepackage{hyperref}
\usepackage{graphicx}
\usepackage[caption=false]{subfig}
\usepackage{multirow}
\usepackage{lmodern}
\usepackage{mathtools}
\usepackage[dvipsnames]{xcolor}
\definecolor{accentRed}{RGB}{192,57,43}
\definecolor{accentBlue}{RGB}{41,128,185}
\usepackage{tikz}
\usetikzlibrary{arrows.meta, bending, decorations.markings}

\newcommand{\dt}[0]{\Delta t}
\DeclareMathOperator{\Var}{Var}
\DeclarePairedDelimiter\abs{\lvert}{\rvert}

\graphicspath{{./figs/}}

\hypersetup{
    colorlinks=true,
    linkcolor=BrickRed,
    citecolor=NavyBlue,
    urlcolor=NavyBlue,
}

\begin{abstract}
    Machine learning methods have recently gained increasing interest as possible means of solving critical slowing down in lattice field theory, with particular emphasis on sampling methods using field transformations such as normalizing flows, annealing/out-of-equilibrium methods, and diffusion models. Annealing methods provide appealing polynomial scaling and exactness guarantees, while diffusion models use a simple and efficient score-matching training procedure. This work explores a method combining these benefits, first training an energy-based diffusion model using score matching, then sampling using either stochastic normalizing flows (SNFs) or flow-transformed replica exchange (T-REX). The choice of sampling method can be made after training the diffusion model; the theoretical properties of SNFs and T-REX and their relative trade-offs are discussed.
Numerical results are presented for scalar field theory and the O(3) non-linear sigma model, using a class of local-footprint energy models with efficient gradient and divergence calculations. Sampling from the trained models follows predicted exactness and scaling properties, and significantly improves over sampling methods without a learned component.
\end{abstract}

\begin{document}

\title{Annealed sampling of diffusion models in lattice field theory}
\date{\today}

\author{Gurtej Kanwar\,\orcidlink{0000-0002-4340-4983}}
\email{gkanwar@ed.ac.uk}
\affiliation{Higgs Centre for Theoretical Physics, School of Physics and Astronomy,\\
University of Edinburgh, EH9 3FD Edinburgh, United Kingdom}

\author{Octavio Vega\,\orcidlink{0000-0002-4360-8126}}
\email{octavio5@illinois.edu}
\affiliation{Illinois Center for Advanced Studies of the Universe and Department of Physics,\\
University of Illinois Urbana-Champaign, Urbana, Illinois 61801, USA}

\maketitle

\section{Introduction}
Lattice field theory is a vital tool for studying strongly interacting field theories. The basis of the approach is a discretization of Euclidean spacetime onto a lattice with spacing $a$, providing a fundamentally non-perturbative regulator for ultraviolet divergences---which are cut off by an effective scale $a^{-1}$---while preserving symmetries such as gauge invariance~\cite{Wilson:1974sk}. By performing Monte Carlo calculations of the regulated path integral at a variety of lattice spacings $a$, then extrapolating $a \to 0$, physical information about the continuum theory can be recovered in a controlled and systematically improvable way.
In the study of the Standard Model, lattice Quantum Chromodynamics (QCD) has for example been used to provide competitive determinations of flavor observables~\cite{FLAG:2024oxs}, parton distribution functions~\cite{Constantinou:2026wii}, hadron form factors~\cite{Hagler:2009ni, Alexandrou:2010cm}, and properties of the QCD phase diagram~\cite{endrodi:2024,borsanyi:2026}. A striking example is hadronic contributions to the anomalous magnetic moment ($g-2$) of the muon, for which lattice QCD has been critical in disambiguating between a new physics discovery scenario and predictions consistent with the Standard Model~\cite{Aliberti:2025beg}.

The systematic and statistical uncertainties propagated by the continuum extrapolation commonly dominate the error budget of lattice field theory calculations. It is therefore important to both extend calculations to finer lattice spacings and obtain increased statistical precision across all Monte Carlo estimates. However, the continuum limit by definition corresponds to a second-order critical point of the theory, where all correlation lengths must diverge in units of the lattice spacing. As this limit is approached, traditional Markov chain Monte Carlo (MCMC) methods based on local updates require increasingly many steps to obtain a fixed number of statistically independent samples. This well-known problem of \emph{critical slowing down}~\cite{Wolff:1989wq} presents a fundamental obstacle to gaining better control of physical observables in strongly interacting quantum field theories, such as QCD. Analogous limits are also of interest in the study of statistical systems near criticality, where similar formulations and Monte Carlo methods are commonly applied.

In recent years, there has been growing interest in investigating alternative Monte Carlo sampling strategies based on generative machine learning approaches. Generative models that use stochastic and deterministic field transformations to transport between probability measures have shown promising early successes in effective and unbiased Monte Carlo sampling of lattice field theories~\cite{Cranmer:2023xbe,Kanwar:2024ujc,Wenger:2026sjp}. While many such methods have been proposed, we broadly classify these into the following strategies:
\begin{enumerate}
    \item \textbf{Normalizing flows} use deterministic transformations to map between a trivial distribution and the target theory~\cite{tabak2010,tabak2013,Rezende:2015icml,Papamakarios:2021,Cranmer:2023xbe,Kanwar:2024ujc}.
    \item \textbf{Annealing methods} explicitly fix a path through the space of theories and aim to follow this path using a combination of deterministic and stochastic field transformations, potentially including MCMC steps. Such methods include trivializing maps~\cite{Luscher:2009eq}, stochastic normalizing flows~\cite{Wu:2020SNF}, and out-of-equilibrium methods~\cite{Jarzynski:1997zdf,Neal:1998ais,Caselle:2016wsw,Caselle:2022acb}.
    \item \textbf{Diffusion models} implicitly fix a path through the space of theories, defined by incrementally applying noise to samples from the target theory, then use deterministic or stochastic transformations to approximately reverse this path~\cite{Sohl:2015dnq,Song:2019gen,Ho:2020dns,Song:2021sgm}.
\end{enumerate}
Each of these methods has by now been applied in a variety of lattice field theory contexts, including trivializing gradient flows for gauge theory~\cite{Luscher:2009eq,Bacchio:2022vje}; normalizing flows for scalar field theories and spin models~\cite{Albergo:2019eim,Hackett:2021idh,deHaan:2021erb,Nicoli:2020njz,Caselle:2022acb,Singha:2022icw,Komijani:2023fzy,Gerdes:2022eve,Albandea:2023wgd,Albandea:2023ais,Mate:2024huc,Kanaujia:2024zrq,Kanaujia:2025jru,Ihssen:2025ybn,Singha:2026slc,Hasenfratz:2026mae,Hasenfratz:2026azk}, gauge theories~\cite{Kanwar:2020xzo,Boyda:2020hsi,Singha:2023xxq}, and theories with dynamical fermions~\cite{Albergo:2022qfi,Abbott:2022hkm,Abbott:2022zsh,Abbott:2024kfc}; stochastic normalizing flows and out-of-equilibrium methods for scalar field theories~\cite{Caselle:2022acb,Albergo:2024trn} and gauge theories~\cite{Bulgarelli:2024brv,Bonanno:2025pdp,Bonanno:2026zqy}; and diffusion models for scalar field theories and spin models~\cite{Wang:2023exq,Vega:2025hgz,Tan:2026ibs,Wang:2026,Tomiya:2026tbc,Grinwald:2026dft}, gauge theories~\cite{Lou:2023obs,Zhu:2024kiu,Zhu:2025pmw,Vega:2025hgz,Aarts:2026zzr,Komijani:2026lan,Komijani:2026ows}, and theories with dynamical fermions~\cite{Vega:2026lrm,Tomiya:2026fjk}.

In all of the above approaches, constructing models that more accurately follow the desired transport path leads to improved sampling quality. In the majority of applications, this has been accomplished by using machine learning parameterizations of the transformations used, along with optimization (i.e., ``training'') using methods based on stochastic gradient descent. More accurate models and thus higher quality sampling can be systematically achieved by making machine learning models more flexible (i.e., ``expressive'') and applying more training.

Unfortunately, training time may be limited by computational costs and extending model expressivity in useful directions may not be straightforward. The cost and effectiveness of training methods also vary between generative approaches. Normalizing flows leave the intermediate path taken through the space of distributions unconstrained. While this can potentially allow flexibility, making a learned transport path easier to find, practical loss functions also can suffer from challenging issues; for example, the reverse Kullback-Leibler (KL) divergence~\cite{Kullback:1951zyt} commonly used for self-training suffers from a well-known mode collapse problem~\cite{Nicoli:2021inv,Nicoli:2023qsl,Kanaujia:2024zrq,Kanaujia:2025jru}. Annealing methods instead use either a path KL divergence~\cite{Wu:2020SNF} or related Physics-Inspired Neural Network (PINN) loss functions~\cite{Albergo:2024trn}, which help to avoid mode collapse but involve solving a costly partial differential equation. In contrast, diffusion models are trained by a relatively cheap score-matching procedure, minimizing a mean-squared-error using training data which have been diffused with varying levels of noise~\cite{Song:2019gen,Song:2021sgm}.
This simple training strategy and its complementarity to other methods motivates our focus in this work on such score-based diffusion models.

The ability to remove or control systematic bias is another important consideration.
For normalizing flows, one can construct reweighting factors by evaluating exact ratios between model and target densities, which can either directly reweight the sampled data~\cite{Nicoli:2019gun} or can be used in Metropolis acceptance tests for asymptotic exactness~\cite{Albergo:2019eim}. The effective sample size (ESS) of directly reweighted flow samples decreases exponentially with the order-2 R\'{e}nyi divergence between the distributions. For annealing methods, total reweighting factors can be obtained by accumulating contributions over each intermediate transport step. Notably, either increasing the number of stochastic MCMC steps or increasing the noise in the continuous stochastic transport applied systematically reduces the variance of log-reweighting factors and thus increases sampling efficiency~\cite{Albergo:2024trn,vargas:2024}; the limit of an infinite number of steps (or noise) corresponds to the \emph{quasi-static limit}, which gives perfect sampling.

On the other hand, strategies for unbiased estimates using diffusion models vary significantly. Sampling using a stochastic reverse process is commonly used in image generation applications~\cite{nichol:2021,saharia:2022,rombach:2022,ramesh:2022}, but does not provide the reweighting factors or exactness guarantees needed for scientific applications. By instead treating the learned score as a deterministic normalizing flow model, the model log-probability and reweighting factors can be exactly computed~\cite{Song:2021sgm}.
It is also possible to determine pathwise weights by specifying a reference forward process and computing a ratio between forward and backward probabilities~\cite{Zhang:2025ubb,Chen:2026bqw,Wang:2026}, which provides unbiased reweighting factors while maintaining stochasticity. Finally, Refs.~\cite{Zhu:2024kiu,Zhu:2025pmw} propose a Metropolis-Adjusted Annealed Langevin Algorithm (MAALA) which combines the stochastic reverse process of a trained diffusion model with MCMC updates near the target distribution for asymptotic exactness. Although effective, the former two methods still lead to a maximum sampling efficiency set by the trained model quality---i.e., they lack an analogue of the quasi-static limit---and the latter method relies on standard MCMC convergence guarantees which may be difficult to control when critical slowing down is an issue.

\begin{figure*}
    \centering
\includegraphics[width=0.95\textwidth]{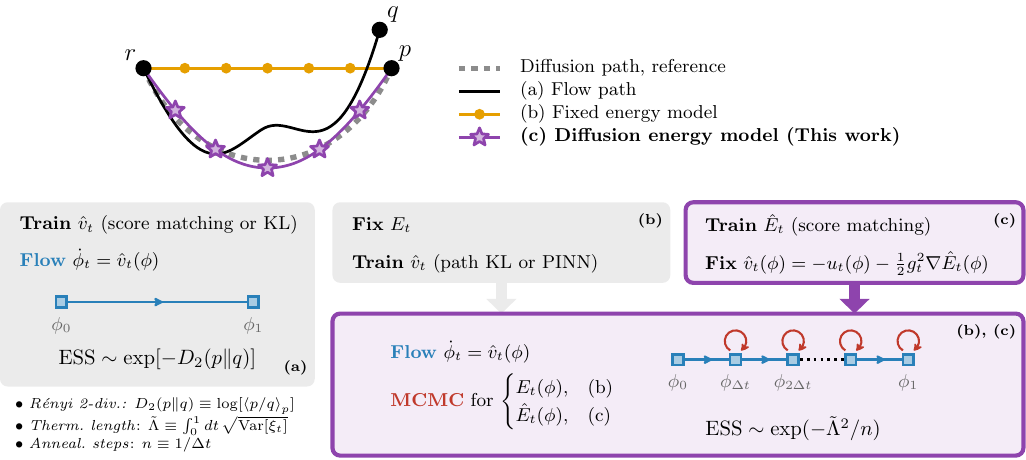}
    \caption{
    Comparison of several strategies to transport between a simple distribution $r$ and a target distribution $p$. Top: Paths taken through the space of distributions by various methods. The reference diffusion path sequentially adds noise to samples from $p$; path (a) is the change-of-density from a continuous field transformation (i.e., a flow); paths (b) and (c) are defined by explicit energy functions evaluated over an annealing time $t \in [0,1]$, where (b) represents an energy function specified \emph{a priori} while (c) represents an energy function trained to match the diffusion path. Bottom: Training and sampling methods used in each case, with the scaling of their respective sampling efficiencies, given in terms of the effective sample size (ESS) per sample drawn. For paths (b) and (c), flow-transformed replica exchange can be used instead of the stochastic normalizing flow shown in the figure; both methods achieve the same scaling, as
detailed in Sec.~\ref{sec:annealed_sampling_methods}.
    }
    \label{fig:summary}
\end{figure*}

In this work, we instead consolidate ideas from annealing methods, diffusion models, and recently developed machine-learning strategies to train and evaluate energy-based diffusion models via annealed sampling, targeting lattice field theory applications. Energy-based diffusion models convert the implicit path fixed by the diffusion process to an explicit path specified by an energy function $\hat{E}_t$, where the energy is trained to approximate the log-probability at each intermediate time $t$, and the endpoints are fixed to match the noise and target distributions.
The learned energy function then enables sampling using standard annealing strategies, which guarantees unbiased expectations and provides a useful scaling limit.
Moreover, the score $-\nabla \hat{E}_t(\phi)$ immediately leads to a field transformation that converges to the optimal one when $\hat{E}_t$ matches the diffusion path. A summary of how this approach relates to various other sampling strategies is visualized in Fig.~\ref{fig:summary}.

Several related ideas have been explored in previous works. Ref.~\cite{Phillips:2024} learns intermediate probability densities within a discrete-time diffusion formulation, using it for an unbiased annealed sampling strategy with particle resampling (Sequential Monte Carlo~\cite{Del:2006smc}); systematic bounds on the variance are established in the limit of an infinite number of steps. Ref.~\cite{Ou:2026sdd} extends the method to systems with discrete state spaces, while Ref.~\cite{Wu:2026rdsmc} develops a training-free approach to estimating the intermediate probability densities. Refs.~\cite{Mate:2024nti,Mate:2025nfe} introduce ``neural thermodynamic integration'', which also learns an interpolating energy function from a diffusion model; the resulting model is used for improved, unbiased partition function estimates, rather than improved sampling and expectation values of microscopic observables. Ref.~\cite{Lee:2026fed} develops a generalized version of free energy and partition function estimation using a learned interpolating energy for more general stochastic processes. Finally, Ref.~\cite{Sexty:2026aym} implements an analogous approach for lattice field theory, demonstrating practical gains in sampling efficiency as the quasi-static limit is approached in the Sequential Monte Carlo framework.\footnote{Ref.~\cite{Sexty:2026aym} was made available during the completion of this manuscript. Our results provide a complementary study using annealed sampling for diffusion models with distinct model architectures, numerical applications, and sampling strategies.}

In the context of this previous work, our key contributions are:
\begin{itemize}
    \item A summary of the annealed sampling method for diffusion models, detailing the theoretical scaling, key metrics to be measured, and subtle issues that must be addressed.
    \item The development and implementation of local-footprint energy models for $\hat{E}_t$, which can be immediately transferred between target theory volumes and support memory-efficient exact divergence calculations, avoiding noisy Hutchinson~\cite{Hutchinson:1989ste,Skilling1989} or spherical-noise estimates~\cite{Peng:2025fpa,Peng:2026fpp}.
    \item A numerical comparison of the stochastic normalizing flow (SNF) and flow-transformed replica exchange (T-REX) methods of sampling from diffusion models, including measured scaling of the sampling efficiency in both approaches; demonstrations are carried out for scalar field theory and the ${\rm O}(3)$ non-linear sigma model.
    \item The implementation of an optimal annealing schedule, demonstrating practical gains that can be achieved by this simple modification.
\end{itemize}

In the remainder of this work, we summarize the annealed sampling method for diffusion models (Section~\ref{sec:method}), discuss the design, implementation, and training of our energy models (Section~\ref{sec:ml-models}), detail our numerical applications (Section~\ref{sec:applications}), discuss the results and implications (Section~\ref{sec:discussion}), and finally conclude with an outlook on future work and impacts (Section~\ref{sec:conclusion}).

\section{Annealed sampling from diffusion models}
\label{sec:method}
We begin with a review in the field theory language of two sequential annealing methods---annealed importance sampling (AIS) and stochastic normalizing flows (SNFs)---which are related by respectively including or excluding a deterministic field transformation. In particular, we introduce the quasi-static limit achieved by taking the number of steps $n$ to infinity, for which the variance of log-reweighting factors asymptotically decreases as $1/n$. These methods are then related to replica exchange methods---ordinary replica exchange (REX)~\cite{Swendsen:1986,Earl:2005} and transformed replica exchange (T-REX)~\cite{Invernizzi:2022,Abbott:2024mix}---which respectively correspond to AIS and SNF by a ``computational duality'' exchanging sequential steps with parallelism~\cite{Syed:2024}. Standard score-based diffusion models are then reviewed using a common language. This background material establishes the basis for the presentation in the remainder of this section of the proposed method.

We use field theory notation as a common language throughout this work, though the conceptual framework extends beyond lattice field theory. Samples will be assumed to be field configurations, generically denoted by $\phi$, where dependence on spacetime sites and internal degrees of freedom has been suppressed. Where necessary, these are made explicit as $\phi^a(x)$, where $a$ indicates an index over internal degrees of freedom and $x \in \mathcal{X}$ denotes a lattice site in the spacetime lattice $\mathcal{X}$. For simplicity, we assume a hypercubic lattice with an extent of $L$ sites in each of $d$ spacetime dimensions, such that $\mathcal{X} \cong (\mathbb{Z}_{L})^{\times d}$ with lattice-units volume $V \equiv |\mathcal{X}| = L^d$. The target distribution to be sampled will be written as
\begin{equation}
    p(\phi) \equiv \frac{1}{Z} e^{-S(\phi)}, \quad Z \equiv \int \mathcal{D}\phi \, e^{-S(\phi)} ,
\end{equation}
and corresponding expectation values for operators $\mathcal{O}$ will be denoted by
\begin{equation}
    \langle \mathcal{O}\rangle_p = \int \mathcal{D}\phi \, \mathcal{O}(\phi) p(\phi) .
\end{equation}
An approximate importance sampling distribution, e.g.\ the sampling distribution of a generative model, will be denoted by the probability density $q(\phi)$.
In this case, appropriate reweighting factors $w(\phi)$ may be used to provide unbiased expectation values of observables as $\langle  w\mathcal{O} \rangle_q = \langle \mathcal{O} \rangle_p$.

\begin{figure}
    \centering
    \begin{tikzpicture}
\draw[-{Stealth[left,round,width=3mm]}, NavyBlue, line width=1pt]
        (0.23,0.05) --
            node[pos=0.0, above=2pt, font=\scriptsize] {$0$}
            node[pos=0.50, above=2pt, font=\footnotesize]
                {Diffusion time $\lambda$}
            node[pos=1.0, above=2pt, font=\scriptsize] {$1$}
        (3.77,0.05);
    \draw[{Stealth[left,round,width=3mm]}-, BrickRed, line width=1pt]
        (0.23,-0.05) --
            node[pos=0.0, below=2pt, font=\scriptsize] {$1$}
            node[pos=0.50, below=2pt, font=\footnotesize]
                {Annealing time $t$}
            node[pos=1.0, below=2pt, font=\scriptsize] {$0$}
        (3.77,-0.05);

    \node[circle, minimum size=8pt, inner sep=0pt, line width=0pt,
          fill=black] (p) at (0,0) {};
    \node[circle, minimum size=8pt, inner sep=0pt, line width=0pt,
          fill=black] (r) at (4.00,0) {};
    \node[circle, minimum size=6pt, inner sep=0pt, line width=2pt,
          draw=white, fill=black] at (1,0) {};
    \node[circle, minimum size=6pt, inner sep=0pt, line width=2pt,
          draw=white, fill=black] at (2,0) {};
    \node[circle, minimum size=6pt, inner sep=0pt, line width=2pt,
          draw=white, fill=black] at (3,0) {};
    \node[anchor=east, font=\footnotesize] at (p.west)
        {$p(\phi)$};
    \node[anchor=west, font=\footnotesize] at (r.east)
        {$r(\phi)$};
\end{tikzpicture}
     \caption{The diffusion time $\lambda$ and annealing time $t = 1-\lambda$ interpolate between the target distribution $\rho_1(\phi) \equiv p(\phi)$ and the noise distribution $\rho_0(\phi) \equiv r(\phi)$.}
    \label{fig:diffusion-time}
\end{figure}
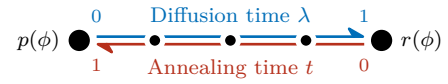

Continuous-time annealed sampling methods and diffusion models both introduce the notion of a continuously evolving probability density as a function of a fictitious annealing or diffusion time. In order to establish common conventions, we denote the (forward) \emph{diffusion time} by $\lambda \in [0,1]$ and the \emph{annealing time}, or reverse diffusion time, by $t \equiv 1-\lambda \in [0,1]$. In all considered methods, the evolving probability density denoted by $\rho_t(\phi)$ corresponds to the target at one endpoint, $\rho_1(\phi) = p(\phi)$, and corresponds to a simple noise distribution at the other endpoint, $\rho_0(\phi) \equiv r(\phi) \equiv \exp(-A(\phi))$.\footnote{In practice, diffusion for a finite interval of time reaches a non-trivial distribution with some dependence on the initial theory. This subtlety is addressed below.}
A schematic depiction of this setup is given in Fig.~\ref{fig:diffusion-time}.

\subsection{Annealed sampling methods}\label{sec:annealed_sampling_methods}
The starting point of the annealed sampling methods discussed in this section is the sequence of distributions
\begin{equation}
\begin{aligned}
    \rho_t(\phi) &\equiv \exp(-E_t(\phi)) / Z_t, \\
    \quad Z_t &\equiv \int \mathcal{D}\phi \, e^{-E_t(\phi)},
\end{aligned}
\end{equation}
which in this section we take to be specified \emph{a priori}. A simple example of such a sequence is a linear interpolation between the target and noise distributions,
\begin{equation}
    E_t(\phi) = t S(\phi) + (1-t) A(\phi).
\end{equation}
The basic idea is to connect the probability distribution $\rho_0$ to the distribution $\rho_1$ by incrementally moving through the intermediate distributions $\rho_t$. For the sequential annealing methods discussed (AIS and SNF), this is done by applying a sequence of reweighting, field transformation, and MCMC steps over time. For the replica-exchange methods discussed (REX and T-REX), this is accomplished by running a parallel set of ``replica'' Markov chains respectively targeting distributions at annealing times $t \in [0,1]$, using swaps between neighboring replicas to improve mixing.

\paragraph{AIS and SNF.}
Sequential annealing methods divide the transport between $\rho_0$ and $\rho_1$ into $n \equiv 1/\dt$ steps, such that samples and weights are iteratively transformed through the sequence
\begin{equation} \label{eq:seq-sampling}
\begin{aligned}
    &&\phi_0 &\;\;\to\;\; \phi_{\dt} &&\to &&\dots& \to& \;\; \phi_1 && \\
   && w_0 &\;\;\to\;\; w_{\dt} &&\to &&\dots &\to& \;\; w_1. &&
\end{aligned}
\end{equation}
Transformations are built such that sampled pairs $(\phi_t,w_t)$ give unbiased expectation values with respect to the distribution $\rho_t$ at each intermediate time $t$,
\begin{equation} \label{eq:rw-expt}
    \langle \mathcal{O} \rangle_{\rho_t} = \frac{\mathbb{E}[w_t \mathcal{O}(\phi_t)]}{\mathbb{E}[w_t]}.
\end{equation}
Here $\mathbb{E}[\cdot]$ indicates the statistical expectation value with respect to possible realizations of this process, typically estimated by averaging over an ensemble sampled according to Eq.~\eqref{eq:seq-sampling}.
In the case of AIS, each transition $(t-\dt) \to t$ consists of two steps~\cite{Neal:1998ais}:
\begin{enumerate}
    \item \textbf{Reweight}: The reweighting factors are updated dynamically using the increment
    \begin{equation} \label{eq:anneal-weights}
    \begin{aligned}
        \Delta \log w_t &\equiv 
        \log w_{t} - \log w_{t-\dt} \\
        &= - E_{t}(\phi_{t-\dt}) + E_{t-\dt}(\phi_{t-\dt}).
    \end{aligned}
    \end{equation}
    The field configuration for the next timestep is initialized to $\phi'_{t} = \phi_{t-\dt}$.
    \item \textbf{MCMC}: Samples are updated using a Markov kernel $\mathcal{K}_{t}$ that satisfies balance with respect to $\rho_{t}$, giving the final field configuration at the next timestep $\phi_{t} \sim \mathcal{K}_{t}(\cdot | \phi'_{t})$.
\end{enumerate}

The latter step is not necessary for correctness, but is important for reducing the variance of reweighting factors. In particular, applying Eq.~\eqref{eq:anneal-weights} alone corresponds to reweighting samples drawn from $\rho_0$ directly to the target $\rho_1$. This is unbiased, but generally results in log-reweighting factors with a variance which scales extensively with the lattice-units volume of the system, $\Var[\log w_1] \sim V$. 
Expectation values as in Eq.~\eqref{eq:rw-expt} will then have exponentially severe noise as $V$ is taken large.

By including a Markov update between each reweighting step, the increments in the log-reweighting factors instead become decorrelated from one another. In the context of lattice field theory, the majority of the fluctuations in the log-weight increments can be expected to be dominated by the short-distance, ultraviolet behavior. This fortunately suggests that the log-weight increments rapidly decorrelate, even with potentially inefficient or locally updating Markov kernels. Assuming perfect local decorrelation---i.e., the short-range fluctuations dominate the log-weights and the Markov kernels resample these fluctuations perfectly---the variance of the total log-reweighting factor splits according to
\begin{equation}
    \Var[\log w_1] = \sum_{i=1}^n \Var[\Delta \log w_{i \dt}].
\end{equation}

When the time step $\dt$ is much smaller than the rate of change of $E_t$, including MCMC updates also guarantees that the sampled distribution is quasi-static, remaining in equilibrium with the distribution $\rho_t$ throughout. Taking $\dt \to 0$ ($n \to \infty$), the Central Limit Theorem then applies for the log-weight increments,
and the asymptotic scaling can be written as~\cite{Syed:2024}
\begin{equation} \label{eq:ais-var}
    \Var[\log w_1] = \frac{1}{n} \int_0^1 dt \, \Var[\partial_t E_t] + O(1/n^2).
\end{equation}
The variance in the integrand is to be computed with respect to the equilibrium distribution $\rho_t$. We denote the integral itself by
\begin{equation}\label{eq:energy-ais}
    \mathcal{S} \equiv \int_0^1 dt \, \Var[\partial_t E_t],
\end{equation}
which is a finite value qualitatively describing the work that must be done to transport along the given path between $\rho_0$ and $\rho_1$. In the lattice field theory context, it carries the same volume dependence identified above, $\mathcal{S} \sim V$. The prefactor $1/n$ in Eq.~\eqref{eq:ais-var} then guarantees that the variance of reweighting factors converges to zero in the quasi-static limit. Moreover, it is only necessary to scale $n \sim V$ to maintain a fixed reweighting variance, in contrast to the exponentially large ensembles that would be required if only importance sampling and direct reweighting were applied~\cite{Neal:1998ais}.

Moving from AIS to SNF sampling requires introducing deterministic field transformations. Applying a generic diffeomorphism $f: \eta \mapsto \phi$ to samples drawn from any initial distribution $\pi^{(I)}$ transforms that distribution via the change-of-variables formula to a new distribution $\pi^{(F)}$ with log-density
\begin{equation}
    \log \pi^{(F)}(\phi) = \log \pi^{(I)}(\eta) - \log \left| \det J_f \right|.
\end{equation}
Here $J_f$ is the Jacobian of the transformation $f$ evaluated at $\eta$.
A parameterized and optimized transformation of this form is a normalizing flow. This leads to a useful improved transport strategy when one is interested in a target distribution $\rho^{(F)}$: use an optimized flow to map $\eta \mapsto \phi$ which follows the importance sampling distribution $\pi^{(F)} \approx \rho^{(F)}$, then apply the exact log-reweighting factors
\begin{equation}
    \log w = \log \rho^{(F)}(\phi) - \log \pi^{(I)}(\eta) + \log \abs{\det J_f}.
\end{equation}
In the limit of a perfect flow, the output samples exactly follow the distribution $\rho^{(F)}$ and the reweighting factors become constant.

In the SNF approach, this concept is used to improve the efficiency of the reweighting step in AIS, applying for each update $(t-\dt) \to t$ the following two steps:
\begin{enumerate}
    \item \textbf{Flow + Reweight}: Transform samples and update the reweighting factors according to
    \begin{equation} \label{eq:snf-flow}
        \phi'_t = f(\phi_{t-\dt}; t,t-\dt),
    \end{equation}
    \begin{equation} \label{eq:snf-weights}
    \begin{aligned}
        \Delta \log w_t &= - E_{t}(\phi'_{t}) + E_{t-\dt}(\phi_{t-\dt})\\
        &\quad\; + \log \abs{\det J_f}.
    \end{aligned}
    \end{equation}
    \item \textbf{MCMC}: Apply the same Markov kernel update as for AIS, $\phi_{t} \sim \mathcal{K}_{t}(\cdot | \phi'_{t})$.
\end{enumerate}
Using the identity transformation $f = \mathrm{Id}$ recovers the AIS procedure, while optimizing the flow transformation $f_t$ at each step strictly improves over this case. In the limit of perfect flow transformations, the reweighting factors become constant and $\Var[\log w_1] = 0$.

To further analyze the scaling of the variance, it is helpful to consider a continuous-time parameterization of the flow transformation:\footnote{For compact manifolds, the integral and sum defining the flow transformation must be taken in the appropriate space, e.g.\ using the exponential map.}
\begin{equation}\label{eq:flow-dt}
\begin{aligned}
    f(\phi_{t-\dt}; t, t-\dt) &= \int_{t-\dt}^t ds \, \hat{v}_s(\phi_s) + \phi_{t-\dt} \\
    \log |\det J_f| &= \int_{t-\dt}^t ds \, [ \nabla \cdot \hat{v}_s(\phi_s)].
\end{aligned}
\end{equation}
Such a definition allows the incremental step size $\dt$ (and the number of steps $n$) to be varied while fixing a particular model of the flow transformation. An analogous quasi-static limit emerges in the limit of $n \to \infty$, in which case the variance converges to the continuous time result~\cite{Syed:2024}
\begin{equation}\label{eq:var_logw_integral_xi}
\begin{aligned}
    \Var[\log w_1] &= \frac{1}{n} \int_0^1 dt \, \Var[\xi_t] + O(1/n^2), \\
    \xi_t &\equiv \partial_t E_t + \hat{v}_t \cdot \nabla E_t - \nabla \cdot \hat{v}_t.
\end{aligned}
\end{equation}
We denote the corresponding integral by
\begin{equation} \label{eq:energy-snf}
    \tilde{\mathcal{S}} = \int_0^1 dt \, \Var[\xi_t].
\end{equation}
An optimal continuous flow $v_t(\phi)$ is one that solves the corresponding transport equation
\begin{equation} \label{eq:transport}
    \nabla \cdot v_t - v_t \cdot \nabla E_t = \partial_t E_t - \langle \partial_t E_t \rangle_{\rho_t},
\end{equation}
which sets the variance and $\tilde{\mathcal{S}}$ to zero~\cite{Albergo:2024trn}. This adds a useful direction for improvement as model complexity and training time can be increased to drive $\hat{v}_t$ towards $v_t$, such that $\tilde{\mathcal{S}}$ is made small.

\paragraph{REX and T-REX.}
Replica exchange methods aim to solve a similar problem to AIS and SNF. Rather than annealing over target distributions in time, the REX and T-REX approaches instantiate $n+1$ parallel replicas respectively targeting the intermediate distributions $\rho_0, \rho_{\dt}, \dots, \rho_1$. We start by defining the more general T-REX algorithm, which can be described by iteratively applying two steps~\cite{Invernizzi:2022,Abbott:2024mix,Zhang:2026apt}:
\begin{enumerate}
    \item \textbf{MCMC:} Update the current sample in each chain with a Markov kernel satisfying balance for its respective target distribution $\rho_t$.
    \item \textbf{Flowed Swap:} For a pair of neighboring chains at annealing time $t$ and $t-\dt$, construct the flowed swap proposal
    \begin{equation} \label{eq:flow-swap}
    \begin{aligned}
        \phi'_t &\equiv f(\phi_{t-\dt}; t, t-\dt), \\
        \phi'_{t-\dt} &\equiv f^{-1}(\phi_t; t, t-\dt).
    \end{aligned}
    \end{equation}
    Accept the swap proposal according to the Metropolis acceptance probability
    \begin{align}
        &p_{\rm swap}(\phi'_t, \phi'_{t-\dt}) = \min(1, w_{\rm swap}(\phi'_t, \phi'_{t-\dt})), \nonumber \\
        &w_{\rm swap}(\phi'_t, \phi'_{t-\dt}) = \label{eq:p-swap} \\
        &\qquad \frac{\rho_t(\phi'_{t}) \rho_{t-\dt}(\phi'_{t-\dt})}{\rho_t(\phi_t) \rho_{t-\dt}(\phi_{t-\dt})}
         \frac{\abs{\det J_f(\phi_{t-\dt})}}{\abs{\det J_f(\phi'_{t-\dt})}}.
        \nonumber
    \end{align}
\end{enumerate}
Throughout this work, we use a non-reversible even-odd swap proposal scheme, alternately proposing swaps between chains $(2j)\dt$ and $(2j+1)\dt$ or between chains $(2j)\dt$ and $(2j-1)\dt$~\cite{Okabe:2001deo}. The ordinary REX algorithm is recovered by using the identity transformation $f = \mathrm{Id}$, for which the proposal is $\phi'_t = \phi_{t-\dt}$ and $\phi'_{t-\dt} = \phi_t$.

\begin{figure*}
    \centering
    \usetikzlibrary{arrows.meta,decorations.pathreplacing,backgrounds}

\definecolor{cold}{named}{NavyBlue}
\definecolor{hot}{named}{BrickRed}

\tikzset{
  state/.style = {circle, minimum size=3mm, inner sep=0pt, draw=white, line width=0.8pt},
  flow/.style  = {-{Stealth[length=2mm]}, line width=0.7pt, gray!70},
  dots/.style  = {loosely dotted, line width=1pt, gray!70, shorten <= 3pt, shorten >= 4pt},
  ghost/.style = {line width=0.6pt, gray!45, dotted, shorten <=1pt, shorten >=1pt},
xflow/.style = {flow, preaction={draw=gray!5, line width=3pt, -}},
  brace/.style = {decorate, decoration={brace,amplitude=4pt}, gray!80, line width=0.6pt},
  lab/.style   = {font=\footnotesize, text=black!70},
  axis/.style  = {-{Stealth[length=2.2mm]}, line width=0.8pt, black!60},
  panel/.style = {rounded corners=6pt, fill=gray!5, draw=gray!25},
}

\begin{tikzpicture}[x=1.1cm, y=0.9cm]
\def\pad{10pt} \def\sfact{0.8} \def\rfact{0.7}

\begin{scope}[local bounding box=L]
\node[font=\bfseries,align=center] at (1.5,0.55) {AIS / SNF};
\node[lab] at (0,-0.3) {$E_0$};
\node[lab] at (1,-0.3) {$E_{\Delta t}$};
\foreach \k in {2,...,3} \node[lab] at (\k,-0.3) {$E_{\k \Delta t}$};
\foreach \i in {1,...,4}{
\foreach \k in {0,...,3}{
    \pgfmathtruncatemacro\p{100-100*\k/3}
    \node[state, fill=cold!\p!hot] (a\i\k) at (\k,-\sfact*\i) {};
  }
  \coordinate (a\i4) at (4,-\sfact*\i);
  \foreach \k/\kn in {0/1,1/2,2/3} \draw[flow] (a\i\k) -- (a\i\kn);
  \draw[dots] (a\i3) -- (a\i4);
  \node[lab] at (4.5,-\sfact*\i) {$\phi^{(\i)},w^{(\i)}$};
}
\draw[brace] (-0.4,-4*\sfact-0.3) -- (-0.4,-1*\sfact+0.3)
  node[midway, left=6pt, lab, align=center] {Parallel\\Walkers};
\draw[axis] (-0.2,-5*\sfact+0.15) -- (3.2,-5*\sfact+0.15)
node[midway, below, lab] {Time};

\end{scope}

\begin{scope}[local bounding box=R]
\node[font=\bfseries,align=center] at (9.5,0.55) {REX / T-REX};
\foreach \t in {1,...,4} \node[lab] at (7+\t,-0.3) 
{$\phi^{(\t)}$};
\node[lab] at (7.35,-\rfact) {$E_{0}$};
\node[lab] at (7.35,-2*\rfact) {$E_{\Delta t}$};
\foreach \k in {2,...,3}{
  \node[lab] at (7.35,-\rfact*\k-\rfact) {$E_{\k \Delta t}$};
}
\foreach \k in {0,...,3}{
  \pgfmathtruncatemacro\p{100-100*\k/3}
  \foreach \t in {1,...,4}
    \node[state, fill=cold!\p!hot] (r\k\t) at (7+\t,-\rfact*\k-\rfact) {};
}
\foreach \t/\k in {2/0, 2/3}{
  \pgfmathtruncatemacro\tn{\t+1}
  \draw[flow] (r\k\t) -- (r\k\tn);
}
\foreach \t/\k in {1/0, 1/2, 2/1, 3/0, 3/2}{
  \pgfmathtruncatemacro\tn{\t+1}
  \pgfmathtruncatemacro\kn{\k+1}
  \draw[ghost] (r\k\t)  -- (r\k\tn);
  \draw[ghost] (r\kn\t) -- (r\kn\tn);
  \draw[flow]  (r\k\t)  -- (r\kn\tn);
  \draw[xflow] (r\kn\t) -- (r\k\tn);
}
\foreach \t in {1,...,4} \draw[dots] (r3\t) -- (7+\t,-5*\rfact-0.2);
\draw[brace] (11.45,-1*\rfact+0.3) -- (11.45,-5*\rfact)
  node[midway, right=6pt, lab, align=center] {Parallel\\Replicas};
\draw[axis] (7.8,-5*\sfact+0.15) -- (11.2,-5*\sfact+0.15)
node[midway, below, lab] {Time};
\end{scope}

\end{tikzpicture}     \caption{Comparison of sequential annealing methods (AIS/SNF) with replica exchange methods (REX/T-REX), demonstrating their relationship through a computational duality that exchanges parallelism over computing resources with sequential evolution in time~\cite{Syed:2024}. The total $n = 1/\Delta t$ steps in annealing time respectively sets the runtime for each AIS/SNF sample and the number of parallel chains for REX/T-REX. The desired total ensemble size (four shown) respectively sets the number of parallel instances of AIS/SNF to be run or the number of steps to be taken in REX/T-REX.}
    \label{fig:comp-dual}
\end{figure*}
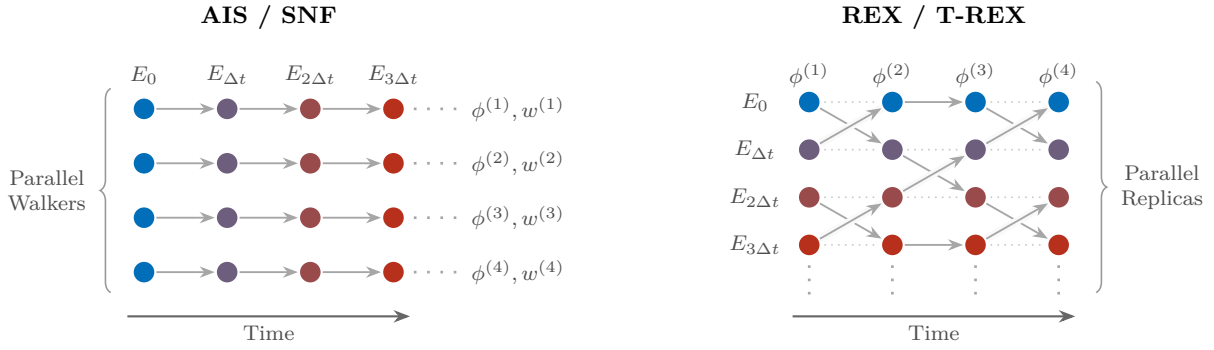

The connection from AIS and SNF sampling to REX and T-REX sampling is given by a computational duality~\cite{Syed:2024}. In AIS and SNF, annealing happens sequentially in time, while statistics are generated in parallel over multiple samples; in REX and T-REX annealing time is distributed over parallel chains, while statistics are generated by evolving the chains sequentially. This relationship is schematically illustrated in Fig.~\ref{fig:comp-dual}.

The result of this duality is that the structure and properties of AIS and SNF largely translate to the case of REX and T-REX. Assuming again a perfect decorrelation of short-distance fluctuations, the efficiency of REX and T-REX is governed by the decorrelation of large scale structures set by the round-trip time $\tau_{\rm RT}$ between the target and noise distributions~\cite{Katzgraber:2006fopt,Lingenheil:2009pte}. The round-trip time can be measured in practice by tracking identities of individual samples as they are swapped through the replicas. We also define the \emph{refresh time}
\begin{equation}
    \tau \equiv \frac{\tau_{\rm RT}}{2(n+1)}.
\end{equation}
The refresh time captures the average number of steps between one new sample being transported from the noise distribution to the target distribution.
Following Ref.~\cite{Syed:2021nrpt}, the equilibrium refresh time with even-odd swap proposals can be exactly related to the average acceptance rates $\mathcal{A}(t, t-\dt) = \mathbb{E}[p_{\rm swap}(\phi'_t, \phi'_{t-\dt})]$ by
\begin{equation}
    \tau = 1 + \sum_{i=1}^n \left( \frac{1}{\mathcal{A}(i\dt,(i-1)\dt)} - 1 \right),
\end{equation}
and it correlates well with the integrated autocorrelation time, $\tau \approx \tau_{\rm int}$.

As $n \to \infty$, the refresh time for the non-reversible swap schedule similarly converges to a limit described by continuous-time quantities~\cite{Syed:2021nrpt,Zhang:2026apt},
\begin{equation}
\begin{aligned}
    \label{eq:tau-rex}
    \tau^{\text{(REX)}} &\to 1 + \frac{1}{2} \int_0^1 dt \, \mathbb{E}\left[|\partial_t E_t(\phi) - \partial_t E_t(\phi')|\right],  \\
    \tau^{\text{(T-REX)}} &\to 1 + \frac{1}{2} \int_0^1 dt \, \mathbb{E}[|\xi_t(\phi) - \xi_t(\phi')|],
\end{aligned}
\end{equation}
where expectation values are taken over $\phi, \phi'$ sampled independently from $\rho_t$.
These expressions simplify further if one assumes a Gaussian distribution for the fluctuations of $E_t(\phi)$ and $\xi_t(\phi)$, which can be expected to hold as the number of lattice sites is taken large in lattice field theory.
The connection can be made explicit by first using a small-$\dt$ expansion of the weights,
\begin{equation}\label{eq:small_dt_expansion_var_logw}
\begin{aligned}
    \chi_t &\equiv \tfrac{1}{2} \Var\left[\log w_{\rm swap}(\phi'_t, \phi'_{t-\dt})\right] \\
    &= -\mathbb{E}\left[\log w_{\rm swap}(\phi'_t, \phi'_{t-\dt})\right] + \text{h.o.} \\
    &= \dt^2 \Var[\xi_t] + \text{h.o.},
\end{aligned}
\end{equation}
where $\xi_t$ reduces to $\partial_t E_t$ if ordinary REX is applied.
Invoking the Gaussian assumption ($\star$) then connects $\chi_t$ to the average acceptance rate as 
\begin{equation}
    \mathcal{A}(t, t-\dt) \stackrel{\star}{=} 1 - \sqrt{\chi_t/\pi} + \text{h.o.},
\end{equation}
and from this to the true asymptotic quantities in Eq.~\eqref{eq:tau-rex}. In the limit of small $\dt$, the Gaussian approximation leads to
\begin{equation}
\begin{aligned}
    \label{eq:tau-rex-gaussian}
    \tau^{\text{(REX)}} &\stackrel{\star}{\to} 1 + \tfrac{1}{\sqrt{\pi}} \Lambda, \quad \Lambda \equiv \int_0^1 dt \, \sqrt{\Var[\partial_t E_t]} \\
    \tau^{\text{(T-REX)}} &\stackrel{\star}{\to} 1 + \tfrac{1}{\sqrt{\pi}} \tilde{\Lambda}, \quad \tilde{\Lambda} \equiv \int_0^1 dt \, \sqrt{\Var[\xi_t]}.
\end{aligned}
\end{equation}
These asymptotic expressions are notably similar to Eqs.~\eqref{eq:energy-ais} and \eqref{eq:energy-snf}. The deeper connection and its relation to choosing an optimal annealing-time schedule is discussed below. We note for now that this makes $\chi_t$ in Eq.~\eqref{eq:small_dt_expansion_var_logw} a useful metric to be measured in the REX/T-REX setting, as it closely relates the sampling performance to that of AIS/SNF.

It has been shown that the constant asymptotic expressions in Eq.~\eqref{eq:tau-rex} hold respectively when $n \gg \Lambda$ or $n \gg \tilde{\Lambda}$~\cite{Zhang:2026apt}. If one is only interested in samples from the target chain, increasing $n$ only increases costs without further benefits after this saturation regime is reached. This motivates a practical compromise in which $n$ is taken just large enough to approach the lower bound efficiently. While $\tau$ cannot then be driven to zero, it is straightforward to build uncorrelated ensembles by keeping only the samples every $\tau$ steps. In this case, the extensivity of the energy $E_t$ implies a scaling with volume as $\Lambda \sim \tilde{\Lambda} \sim \sqrt{V}$, and the cost of the process in units of the number of MCMC or flow steps should scale as $n \, \tau \sim V$ to achieve approximately a fixed number of independent samples. This is equivalent to the scaling expected from sequential annealing methods.

We conclude this discussion by noting that Eq.~\eqref{eq:tau-rex} ensures that a flow $v_t$ solving the transport equation in Eq.~\eqref{eq:transport} is optimal for both SNF and for T-REX. The choice of sampling strategy can therefore be made \emph{after} constructing and training a given flow model, allowing one to exchange time and parallelism to optimally allocate computing resources at sampling time. In this work, we explore both avenues to sampling as possible applications once a diffusion model has been trained using our proposed strategy.

\paragraph{Optimal schedules.}
The connection between the continuous-time metrics for AIS/SNF and REX/T-REX stems from a Fisher-Rao geometry in the space of distributions~\cite{Crooks:2007,Feng:2009}. Of relevance here, this gives meaning to $\Lambda$ and $\tilde{\Lambda}$ as path lengths in this space; one can show that these path lengths are true geometric quantities, independent of parameterization. For $\tilde{\Lambda}$, the inclusion of the deterministic flow $v_t$ acts to modify the metric and (typically) shorten the path length relative to $\Lambda$. A connection can then be made to the quantities $\mathcal{S}$ and $\tilde{\mathcal{S}}$ defined in Eqs.~\eqref{eq:energy-ais} and \eqref{eq:energy-snf}: these are proportional to the action taken along the respective paths, and their action is minimized when the path is traversed at a rate of equal action. This can be achieved by rescheduling the evolution in annealing time $t \to t'$ such that $\Var[\partial_{t'} E_{t'}]$ or $\Var[\xi_{t'}]$ is constant along the entire path, and in this case the optimally scheduled action converges respectively to
\begin{equation}\label{eq:optimal_annealed_action}
\begin{aligned}
    \mathcal{S} &\to \left( \int_0^1 dt' \, \sqrt{\Var[\partial_{t'} E_{t'}]} \right)^2 = \Lambda^2, \\
    \tilde{\mathcal{S}} &\to \left( \int_0^1 dt' \, \sqrt{\Var[\xi_{t'}]} \right)^2 = \tilde{\Lambda}^2.
\end{aligned}
\end{equation}
In general, these are lower bounds, $\mathcal{S} \geq \Lambda^2$ and $\tilde{\mathcal{S}} \geq \tilde{\Lambda}^2$. Choosing the optimal schedule saturates these bounds, giving a constant factor improvement, but does not change the scaling with volume which is set by $\Lambda$ or $\tilde{\Lambda}$ themselves. These are also respectively the optimal choices for REX and T-REX under the Gaussian assumption; in a non-Gaussian setting, the optimal schedule instead involves equalizing the infinitesimal contributions in Eq.~\eqref{eq:tau-rex}, which can be estimated by equalizing swap acceptance rates across all pairs of chains~\cite{Syed:2021nrpt}. We explore a specific example of this in Sec.~\ref{sec:phi4_trex_demo}.

\subsection{Diffusion models}
In contrast to the above annealing methods, diffusion models implicitly fix a path between the target distribution $\rho_1(\phi) \propto \exp(-S(\phi))$ and the noise distribution $\rho_0(\phi) \propto \exp(-A(\phi))$, which is defined by incrementally adding noise from $t=1$ to $t=0$. For convenience, we describe this process in terms of the diffusion time $\lambda = 1-t$. The evolution of individual samples in a generic diffusion model setup can then be described by a Langevin equation evolving forwards in diffusion time as
\begin{equation}\label{eq:forward_ito_sde}
    d \phi_{\lambda} = u_\lambda(\phi_\lambda) d\lambda + g_\lambda dW_\lambda,
\end{equation}
where $u_\lambda(\cdot)$ is an arbitrary vector field, $g_\lambda$ sets the scale of the incremental noise,
and $dW_\lambda$ is Wiener noise normalized such that $\mathbb{E}[dW_\lambda^2] = d\lambda$.
The associated Fokker-Planck equation determines the evolution of the probability density in diffusion time. Using the notation $\rho_\lambda = \exp(-E_\lambda(\phi)) / Z_\lambda$ in analogy to annealing methods, the Fokker-Planck equation on the energy is given by
\begin{equation} \label{eq:fpe-diff-lam}
\begin{aligned}
    \partial_\lambda E_\lambda - \langle \partial_\lambda E_\lambda \rangle &= \nabla \cdot u_\lambda - u_\lambda \cdot \nabla E_\lambda\\
    &\quad + \tfrac{1}{2} g_\lambda^2 (\nabla^2 E_\lambda - \lVert \nabla E_\lambda \rVert^2).
\end{aligned}
\end{equation}
To arrive at the left expression, we use the identity
\begin{equation}
    \partial_\lambda \log Z_\lambda = (\partial_\lambda Z_\lambda) / Z_\lambda = -\langle \partial_\lambda E_\lambda \rangle.
\end{equation}
Rewriting in terms of the annealing time $t$ gives
\begin{equation} \label{eq:fpe-anneal-t}
\begin{aligned}
    \partial_t E_t - \langle \partial_t E_t \rangle
    &= -\nabla \cdot u_t + u_t \cdot \nabla E_t \\
    &\quad - \tfrac{1}{2} g_t^2 \left(\nabla^2 E_t - \lVert \nabla E_t \rVert^2\right),
\end{aligned}
\end{equation}
where we abuse notation to evaluate all relevant quantities $X$ according to the replacement $X_t \equiv X_\lambda|_{\lambda=1-t}$.

The standard sampling strategy for score-based diffusion modeling is to apply a corresponding Langevin evolution in $t$ that reproduces Eq.~\eqref{eq:fpe-anneal-t} in the space of distributions. For example, the commonly considered family of such processes can be defined as
\begin{equation} \label{eq:reverse-sde}
    d\phi_t = -u_t(\phi_t) dt - \tfrac{1}{2}(g_t^2 + \tilde{g}_t^2) \nabla E_t dt + \tilde{g}_t dW_t,
\end{equation}
in terms of an arbitrary additional noise coefficient $\tilde{g}_t$.
Applying this Langevin equation to transport from easily generated noise samples to the target distribution is typically termed ``reverse diffusion'' or ``denoising''.

To apply Eq.~\eqref{eq:reverse-sde}, it is sufficient to estimate or learn the score $-\nabla E_t$. While most machine-learning approaches parameterize and match the score directly, we here focus on energy-based models (EBMs)~\cite{Teh:2003,Du:2020,Salimans:2021ebms}, which construct and train a parameterization $\hat{E}_t(\phi) \approx E_t(\phi) + C_t$. While traditional score-matching aims to match a score network $\hat{s}$ to satisfy $\hat{s}_\lambda \approx -\nabla E_\lambda$, here the gradient of the energy $-\nabla \hat{E}_\lambda$ can be used directly in the same simple regression loss to converge to the true energy function, up to an irrelevant constant. From $\hat{E}_t$, the estimated score $-\nabla \hat{E}_t \approx -\nabla E_t$ can be used directly in Eq.~\eqref{eq:reverse-sde}. The constant $C_t$ encodes an ambiguity in the definition of $E_t$ and $\hat{E}_t$ that is not fixed by standard training procedures, but that also does not affect sampling or measurements.

\subsection{Sampling from EBMs using annealing methods}
\label{sec:diff-anneal-sampling}
In this work we choose to treat the learned energy function $\hat{E}_t$ as a continuous annealing path to be sampled by any of the standard methods outlined earlier in this section. The key observation here is that the true annealing path $E_t$ of the diffusion process automatically leads to an optimal deterministic field transformation defined by
\begin{equation} \label{eq:diff-flow}
    v_t(\phi) = -u_t(\phi) - \tfrac{1}{2} g_t^2 \nabla E_t(\phi).
\end{equation}
In particular, this continuous flow satisfies the transport equation in Eq.~\eqref{eq:transport} exactly because the energy $E_t(\phi)$ satisfies the defining Fokker-Planck equation in Eq.~\eqref{eq:fpe-anneal-t}:
\begin{equation}
\begin{aligned}
    &\partial_t E_t - \langle \partial_t E_t \rangle \\
    &\quad = -\nabla \cdot u_t + u_t \cdot \nabla E_t - \tfrac{1}{2} g_t^2 (\nabla^2 E_t - \lVert \nabla E_t \rVert^2) \\
    &\quad = -\nabla \cdot (u_t + \tfrac{1}{2} g_t^2 \nabla E_t) + (u_t + \tfrac{1}{2} g_t^2 \nabla E_t) \cdot \nabla E_t \\
    &\quad = \nabla \cdot v_t - v_t \cdot \nabla E_t.
\end{aligned}
\end{equation}
A perfectly trained EBM can thus be immediately used to sample via the SNF method with zero reweighting noise or via the T-REX method with perfectly efficient swaps.

Motivated by this connection, we advocate for defining diffusion models using EBMs, training via score matching, and sampling using either the SNF or T-REX method in order to leverage the complementary features of these approaches. This strategy is summarized in Algorithm~\ref{fig:main-algo}.

\begin{algorithm*}[t]
\caption{Training and annealed sampling from an energy-based diffusion model}\label{fig:main-algo}
\SetAlgoLined
\BlankLine
\textbf{Diffusion settings:} Choose a drift $u_\lambda$ and noise schedule $g_\lambda$. \\[0.5ex]
\textbf{Model definition:} Parameterize $\hat{E}_t(\phi)$, mapping $t \in [0,1]$ and a field $\phi$ to a scalar output. The model must fix $\hat{E}_0(\phi) = A(\phi)$ and $\hat{E}_1(\phi) = S(\phi)$. \\[0.5ex]
\textbf{Training:}
\begin{enumerate}
    \item Generate training data $\phi$ drawn from the target distribution $p(\phi) \propto \exp(-S(\phi))$.
    \item Train the model $\hat{E}_t(\phi)$ using score matching.
\end{enumerate}
\textbf{Sampling:}
Choose annealing step size $\dt = 1/n$. Define $\hat{v}_t(\phi) \equiv -u_t(\phi) - \tfrac{1}{2} g_t^2 \nabla \hat{E}_t(\phi)$. Choose either SNF or T-REX. \\[0.5ex]
$\triangleright$ \underline{SNF:}\\
\-\hspace{10pt}For each independent sample to be drawn, apply the following to sample a weighted pair $(\phi_1, w_1)$.
\begin{enumerate}
    \item Sample $\phi_0$ from the noise distribution $r(\phi) \propto \exp(-A(\phi))$. Set $w_0 = 1$.
    \item Per step $(t-\dt) \to t$: apply the flow-reweighting step in Eqs.~\eqref{eq:snf-flow} and \eqref{eq:snf-weights}, mapping $(\phi_{t-\dt},w_{t-\dt}) \to (\phi'_{t},w_t)$; apply a Markov update $\phi'_t \to \phi_t$ satisfying balance with respect to $\hat{E}_t$.
\end{enumerate}
$\triangleright$ \underline{T-REX:}
\begin{enumerate}
    \item Initialize $n+1$ Markov chains corresponding to $\hat{E}_0, \hat{E}_{\dt}, \dots, \hat{E}_1$.
    \item Repeat: apply MCMC updates and propose swaps according to Eqs.~\eqref{eq:flow-dt}, \eqref{eq:flow-swap}, \eqref{eq:p-swap}, using $\hat{v}_t$ for the transport.
\end{enumerate}
\end{algorithm*}

We highlight below four practical issues that may challenge the implementation or convergence towards the optimal solution in Eq.~\eqref{eq:diff-flow}, proposing solutions to each.

The first issue is that only the learned approximation $\hat{E}_t$ is available, rather than the exact energy $E_t$ along the diffusion path. It is then necessary to define the continuous flow in practice by
\begin{equation} \label{eq:diff-flow-real}
    \hat{v}_t(\phi) = -u_t(\phi) - \tfrac{1}{2} g_t^2 \nabla \hat{E}_t(\phi).
\end{equation}
For any finite model trained for a finite time, $\hat{E}_t$ will not exactly satisfy the defining Fokker-Planck equation and $\hat{v}_t$ will not exactly satisfy the transport equation in Eq.~\eqref{eq:transport}. However, the relevant metrics for annealed sampling are driven by $\Var[\xi_t]$, which is the variance of the residual in the transport equation, determined by the score $-\nabla \hat{E}_t$ and subtracted time derivative $\partial_t \hat{E}_t - \langle \partial_t \hat{E}_t \rangle$. The score is directly optimized to match the solution to this equation via score matching; the subtracted time derivative is also constrained as a close approximation $\hat{E}_t(\phi) \approx E_t(\phi) + C_t$ leads to good agreement.
In practice, monitoring the respective SNF or T-REX metrics can be used to validate the sampling efficiency resulting from applying Eq.~\eqref{eq:diff-flow-real}. We note that using an imperfect model does not affect the correctness guarantees inherent in SNF and T-REX applications.

The second issue is that the annealing path to be sampled must exactly match the target distribution and noise distribution at the endpoints. Using a free approximation may lead to deviations $\hat{E}_1(\phi) \neq S(\phi) + C_1$ or $\hat{E}_0(\phi) \neq A(\phi) + C_0$, which either introduces bias or noise from additional reweighting factors which is not removed by scaling the number of annealing steps in the methods above. We therefore construct machine-learning parameterizations $\hat{E}_t$ that automatically enforce the endpoint conditions.

The third issue is that diffusion over the finite interval $\lambda \in [0,1]$ does not exactly converge to the specified noise distribution, $\rho_{\lambda=1}(\phi) \neq r(\phi)$. In particular, the noise distribution $r(\phi)$ is typically only recovered in the $\lambda \to \infty$ limit of the diffusion process. In practice, the diffusion coefficient $g_\lambda$ can be chosen large enough that at diffusion time $\lambda = 1$, the true distribution is a very good approximation of the noise distribution. By then fixing the endpoint exactly as $\hat{E}_0(\phi) = A(\phi)$, this introduces a (likely very mild) tension between the energy model $\hat{E}_t$ and the true distributions under diffusion, limiting the quality of agreement with the defining Fokker-Planck equation in Eq.~\eqref{eq:fpe-anneal-t}. While this theoretically bounds the achievable model quality specified by $\tilde{\mathcal{S}}$ in Eq.~\eqref{eq:energy-snf}, in practice we expect the model parameterization and training time to be more significant constraints.

The fourth issue is a practical implementation challenge. Both the SNF and T-REX methods involve integrating the vector field $\hat{v}_t$ or its divergence over an interval of annealing time $\dt$.
For sufficiently unbiased or asymptotically exact sampling, it is necessary to control the integration error in these steps. In our practical demonstrations, we choose to address this by using high-precision Runge-Kutta (RK4) integrators which converge with a much faster power of $\dt$ than the corresponding statistical uncertainties. For this approach, it is also necessary to have an efficiently computable divergence,
\begin{equation}\label{eq:divergence_of_drift}
    \nabla \cdot \hat{v}_t = - \nabla \cdot u_t - \tfrac{1}{2} g_t^2 \nabla^2 \hat{E}_t.
\end{equation}
Typical choices for the drift $u_t$ are simple and their divergences can be analytically computed; the design constraint is then to ensure that the Laplacian $\nabla^2 \hat{E}_t$ is tractable. In the following section, we introduce a particular architecture with this property and focus on this approach as the primary strategy for our numerical applications.

Finally, we conclude this section by summarizing the main advantages of using annealed sampling with trained diffusion models instead of other possible sampling methods. First, in contrast to sampling the reverse diffusion process, exactness can be guaranteed. Second, applying a sequential annealing method (AIS or SNF) provides a quasi-static limit in which reweighting noise can be driven arbitrarily close to zero. Similarly, replica exchange methods (REX or T-REX) provide an analogous scaling direction with increasing $n$ and update steps.
Third, as opposed to annealed sampling along a fixed annealing path, here the learned path comes with an approximately optimal field transformation $\hat{v}_t$. Finally, the training in this case uses the simple score-matching loss, rather than the more complicated loss functions typically used within flow-based annealing methods to learn a $\hat{v}_t$ suitable for a fixed path~\cite{Wu:2020SNF,Albergo:2024trn}.

\section{Machine learning setup}
\label{sec:ml-models}
In this section we outline our choices in parameterizing energy-based models $\hat{E}_t$ in a way that is suitable for training via score-matching and sampling via annealing methods without introducing additional bias or noise (e.g., by avoiding stochastic trace estimators). We further discuss loss functions and practical training strategies used here, with emphasis on mitigating noise inherent in the score-matching loss.

\subsection{Energy-Based Model Parameterizations}\label{sec:ebm_parameterizations}

Throughout this work, we use an endpoint-constrained energy function given by
\begin{equation} \label{eq:interp-ebm}
    \hat{E}_t(\phi) = a_t A(\phi) + b_t S(\phi) + c_t\hat{f}_t(\phi),
\end{equation}
where $\hat{f}_t(\phi)$ is parameterized by a neural network that maps a time $t$ and field configuration $\phi$ to a scalar value.
The coefficients are chosen so that $\hat{E}_0(\phi) = A(\phi)$ and $\hat{E}_1(\phi) = S(\phi)$, which requires
\begin{align}
    a_0 &= 1, \quad a_1 = 0, \\
    b_0 &= 0, \quad b_1 = 1, \\
    c_0 &= c_1 = 0.
\end{align}
In this work, we always use a simple polynomial interpolation $a_t = (1-t)$, $b_t = t$, and $c_t = t(1-t)$.
This allows us to leave the interior energy function $\hat{f}_t$ unrestricted.

In the following applications, the noise distribution will be fixed analytically to be either the uncorrelated Gaussian distribution defined by $A(\phi) = \sum_x \lVert \phi(x) \rVert^2 / 2$, or a uniform distribution corresponding to $A(\phi) = 0$.
The target action $S(\phi)$ is known analytically for each theory of interest and can be evaluated directly on field configurations.

To build an expressive dependence on time $t$ in the network $\hat{f}_t$, it is beneficial to construct a richer feature representation for the raw times $t$. Following standard practice in score-based generative modeling~\cite{Karras:2022}, we first map the scalar $t \in [0,1]$ to a higher-dimensional feature vector by a separate neural network.
Our models share a common time encoding network, composed as a three-layer fully-connected neural network with SiLU activation functions~\cite{Elfwing:2018}. The output embedded time vector $t_{\rm enc}$ is then composed with dependence on $\phi$ in theory-specific and architecture-specific ways detailed in the applications in Sec.~\ref{sec:applications}.

When designing the dependence on $\phi$ in the network $\hat{f}_t$, we note that both the score and Laplacian of $\hat{E}_t$ are needed for annealed sampling.
Distributing over the components by linearity gives
\begin{equation}
\begin{aligned}
    -\nabla \hat{E}_t &= -a_t \nabla A - b_t \nabla S - c_t \nabla \hat{f}_t, \\
    \nabla^2 \hat{E}_t &= a_t \nabla^2 A + b_t \nabla^2 S + c_t \nabla^2 \hat{f}_t.
\end{aligned}
\end{equation}
For the noise distributions and actions used in this work, both the score and Laplacian are efficiently computed analytically. Either the parameterization $\hat{f}_t$ can be designed to support these operations efficiently, stochastic estimators may be used to estimate these values quickly, or a brute force direct calculation must be used. The latter is impractical for large lattice volumes $V$. Our choices for parameterizing the interior energy $\hat{f}_t$ are detailed below, including computational aspects relevant for the score and Laplacian.

\begin{figure*}
    \centering
    \includegraphics[width=0.9\linewidth]{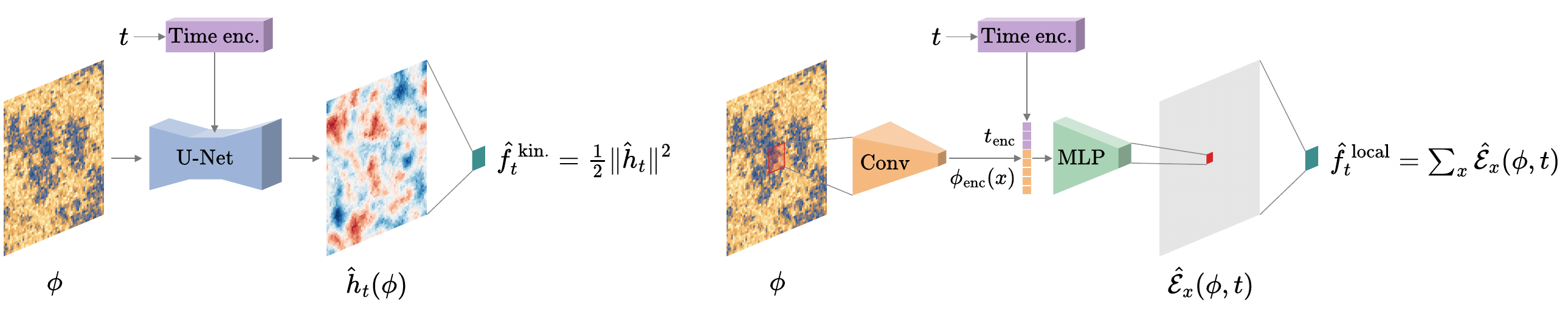}
    \caption{Comparison of kinetic (left) and local (right) EBM parameterizations used in this work.}
    \label{fig:ebm-arch}
\end{figure*}

\paragraph{Kinetic EBM.}
Inspired by the functional form of the kinetic energy function, a common choice of parameterization for EBMs can be written as
\begin{equation}
    \hat{f}_t^{\rm kin.}(\phi) = \frac{1}{2}\left\| \hat{h}_t(\phi) \right\|^2,
\end{equation}
where the vector field $\hat{h}_t(\phi)$ is itself represented by an arbitrary, dimension-preserving neural network. This setup has the benefit of computing a scalar function using the more expressive architectures one would typically use for a vector field, such as a U-Net~\cite{Ronneberger:2015}. Computing the Laplacian of $\hat{f}$, however, can be computationally intensive as such black-box parameterizations require either a brute-force $O(V^2)$ calculation or use noisy stochastic trace estimators. Regardless, we include this architecture in numerical comparisons presented in the following sections.

\paragraph{Local-footprint EBM.}
Another option is to construct a translationally-invariant energy function defined over local patches of the lattice,
\begin{equation} \label{eq:local-footprint-ebm}
    \hat{f}_t^{\text{local}}(\phi) = \sum_x \hat{\mathcal{E}}_x(\phi, t).
\end{equation}
Here each local energy term $\hat{\mathcal{E}}_x(\phi,t)$ is restricted to act on only the degrees of freedom within a $K \times K \times \cdots \times K$ patch centered on $x$; we denote the sites in the patch by
\begin{equation}
    \mathcal{B}_r(x) = \left\{y \in \mathcal{X} : \|x-y\|_\infty\leq r\right\},
\end{equation}
with $r = \lfloor K/2 \rfloor$. The footprint size $K$ provides a natural hyperparameter to trade off expressivity versus cost. This explicitly captures the physical locality of typical operators expected to appear in the calculation of the energy (or action) in lattice field theory. A further practical benefit of this choice is that $\hat{f}^{\rm local}_t$ can be evaluated on volumes different from the one used to train the model by summing.

Unlike a typical convolutional network, the strictly local footprint guarantees that the score $-\nabla \hat{f}_t$ evaluated at site $x$ only depends on fields at most a distance $K$ in each direction from that site, regardless of the complexity of the network $\hat{\mathcal{E}}_x$. This allows the score to be evaluated by a memory-efficient sum over contributions from each local patch, avoiding the need to materialize a computational graph over the entire lattice geometry.
Similarly, the Laplacian can be computed by a local sum,
\begin{equation}
    \nabla^2 \hat{f}_t(\phi) = \sum_x \sum_{y \in \mathcal{B}_r(x)} \nabla_y^2 \hat{\mathcal{E}}_x(\phi, t),
\end{equation}
which requires only $O(V)$ time and volume-independent memory usage.

Our parallelized implementation of this architecture begins by collecting the local degrees of freedom within the $K^d$ sites of each patch to a hidden vector $\phi_{\rm enc}(x)$ at site $x$ using a preprocessing network. In the simplest cases, this preprocessing network is a basic convolution that linearly maps all components within the patch into the hidden vector; for the case of the ${\rm O}(3)$ model considered below, we instead focus on an architecture that builds all invariant inner products within the patch before applying a linear map into the hidden vector.
The hidden state and the encoded time are then concatenated on each site to produce the local state vector $[\phi_{\rm enc}(x), t_{\rm enc}]$. This state is subsequently processed locally in parallel across all sites by a single neural network comprised of a sequence of fully-connected residual blocks. This network outputs the scalar value $\hat{\mathcal{E}}_x(\phi,t)$; finally this state is summed according to Eq.~\eqref{eq:local-footprint-ebm}.

\subsection{Training Diffusion EBMs}
Diffusion EBMs can be trained in largely the same manner as traditional diffusion models according to the score-matching procedure. Since training pertains to the forward process in Eq.~\eqref{eq:forward_ito_sde}, we again use diffusion time $\lambda \in[0, 1]$ for the discussion in this section.

The score-matching objective function is
\begin{equation} \label{eq:score-matching}
    \mathcal{L} = \int_0^1 d\lambda \int \mathcal{D}\phi \, \rho_\lambda(\phi) \left\|\nabla \hat{E}_\lambda(\phi) + \nabla \log \rho_\lambda(\phi)\right\|^2,
\end{equation}
which is minimized when $-\nabla \hat{E}_\lambda$ matches the exact score.
Evaluating Eq.~\eqref{eq:score-matching} requires samples from the intermediate distributions $\rho_\lambda(\phi)$. These can be obtained by drawing $\phi_0$ from the target distribution $\rho_{\lambda=0}$ and evolving to the diffusion time $\lambda$ by applying appropriate random noise, typically sampled directly using the heat kernel $K_\lambda(\phi | \phi_0)$ associated with the diffusion process in Eq.~\eqref{eq:fpe-diff-lam}.
In all cases studied here, $K_\lambda$ only depends on a global noise level denoted by $\sigma_\lambda$; for example, for Euclidean diffusion with or without a linear drift, the sampled kernel is a Gaussian with width $\sigma_\lambda$. For two common choices of the forward drift $u_\lambda$ in Eq.~\eqref{eq:forward_ito_sde}, $\sigma_\lambda$ satisfies
\begin{equation}
    \sigma_\lambda^2 = 
    \begin{cases}
        \int_0^\lambda g_s^2 ds, &u_\lambda(\phi) \equiv 0, \\
        1 - \exp\left[-\int_0^\lambda g_s^2 ds\right], &u_\lambda(\phi) = -\frac{1}{2}g_\lambda^2 \phi.
    \end{cases}
\end{equation}
In this work, the training data $\{ \phi_0 \}$ is sampled online, using standard MCMC updates to refresh a training ensemble periodically. We train using smaller-than-physical volumes so that this generation step requires much lower computational effort than sampling at the target volume.

The theoretical score-matching objective in Eq.~\eqref{eq:score-matching} relies on the true score $\nabla \log \rho_\lambda(\phi)$, which is not analytically known. Instead, it is replaced by an unbiased estimator using one of two methods: denoising score matching (DSM) and target score matching (TSM)~\cite{debortoli:2024}. The methods are respectively based on the denoising score identity (DSI),
\begin{equation} \label{eq:dsi}
    \nabla \log \rho_\lambda(\phi_\lambda) = \mathbb{E}_{\phi_0 | \phi_\lambda}[\nabla_\lambda \log K_\lambda(\phi_\lambda | \phi_0)],
\end{equation}
and the target score identity (TSI), which takes a kernel-dependent form,
\begin{equation} \label{eq:tsi}
    \nabla \log \rho_\lambda(\phi_\lambda)
    = \mathbb{E}_{\phi_0|\phi_\lambda}[\mathcal{T}_{\lambda}(\phi_\lambda, \phi_0)].
\end{equation}
Above, $\nabla_\lambda \equiv \partial/\partial \phi_\lambda$, and the heat kernel $K_\lambda$ matches the one used to apply noise to training data.
For example, in the simplest case of Euclidean space and a symmetric kernel $K_\lambda(\phi_\lambda | \phi_0) \equiv K_\lambda(\phi_\lambda - \phi_0)$, the TSI operator reduces to the score evaluated on the target-distribution samples, $\mathcal{T}_\lambda(\phi_\lambda,\phi_0) = -\nabla_0 S(\phi_0)$.
The extension to a compact Riemannian manifold is treated in the numerical application of Sec.~\ref{sec:o3_model_setup}.

Inserting these definitions gives two equivalent practical loss functions to be optimized,\footnote{A constant which does not affect optimization has been dropped while moving the expectation value outside the squared norm.}
\begin{equation}
\begin{aligned}
    \mathcal{L}^{\rm DSM}_\lambda &= \mathbb{E}\left[\left\| \nabla \hat{E}_{\lambda}(\phi_\lambda) + \nabla_\lambda \log K_\lambda(\phi_\lambda|\phi_0)\right\|^2\right], \\
    \mathcal{L}^{\rm TSM}_\lambda &= \mathbb{E}\left[\left\| \nabla \hat{E}_{\lambda}(\phi_\lambda) + \mathcal{T}_\lambda(\phi_\lambda,\phi_0) \right\|^2\right].
\end{aligned}
\end{equation}
In both cases, we have defined the loss for a fixed value of $\lambda$, and the expectations are taken over $\phi_0$ drawn from the target distribution, with corresponding $\phi_\lambda$ drawn as described above.
Both objective functions are tractable on their own, but their noise properties differ: DSM has low noise as $\lambda$ is taken large, while TSM has low noise as $\lambda \to 0$~\cite{debortoli:2024}. This motivates two practical modifications. First, we follow common practice and rescale the DSM loss according to $\sigma_\lambda^2$ to normalize the noise uniformly over the domain. Second, we use a composite loss, partitioning the diffusion time into two sub-intervals $[0,1] = [0, \lambda_{\rm cut}] \cup[\lambda_{\rm cut}, 1]$, and applying the complementary objectives as one combined loss function,
\begin{equation} \label{eq:composite-loss}
    \mathcal{L} = \int_0^{\lambda_{\rm cut}} d\lambda \, \mathcal{L}^{\rm TSM}_\lambda + \int_{\lambda_{\rm cut}}^1 d\lambda \, \sigma_\lambda^2 \mathcal{L}^{\rm DSM}_\lambda.
\end{equation}
This composite loss is used for all training involved in numerical applications below.

\subsection{MCMC using EBMs}
Both SNF and T-REX involve MCMC updates with respect to the learned energy function $\hat{E}_t(\phi)$. A straightforward Markov chain update in this case is provided by applying Hybrid Monte Carlo~\cite{Duane:1987de,Gupta:1992np,Kennedy:1993jy} using the learned score $-\nabla \hat{E}_t$ as the force term and $\hat{E}_t$ itself for the necessary Metropolis acceptance test. This implementation is used for all MCMC sampling in numerical applications below.

We also note that local Metropolis updates can be applied for local-footprint energy models $\hat{E}_t$ as long as local updates proposed in parallel are separated by a minimum distance of $K$ in each direction. This strategy is not used here, but may be practically beneficial in future work.

\section{Applications}\label{sec:applications}
To test the effectiveness of annealed sampling with energy-based diffusion models for lattice field theory applications, we perform numerical and analytical checks using both scalar field theory and the ${\rm O}(3)$ non-linear sigma model in two dimensions. All lattice calculations use a periodic lattice with $L^d$ sites and unit spacing. The basic physics and diffusion model setups are discussed for each theory in turn.

\subsection{Scalar field theory}
For scalar field theory applications, we use the standard nearest-neighbor Euclidean action,
\begin{gather}
    S(\phi) = \sum_x \Bigg[\tfrac{1}{2}\sum_\mu (\phi(x+\hat\mu) - \phi(x))^2 + V[\phi(x)]\Bigg], \nonumber \\
    V[\phi(x)] = \tfrac{m^2}{2} \phi(x)^2 + \lambda \phi(x)^4.
    \label{eq:lattice_scalar_action}
\end{gather}
With the exception of the targeted numerical study for the $\phi^4$ interacting theory in Sec.~\ref{sec:phi4_trex_demo}, our numerical and theoretical results focus on the free scalar theory, defined by setting $\lambda = 0$. 
In this case, using the unitary discrete Fourier transforms on the lattice, $\phi$ may be written as the inverse Fourier transform of its momentum-space counterpart $\tilde{\phi}$ as
\begin{equation}\label{eq:inv_fourier_transform_discrete}
     \phi(x) = \frac{1}{L^{d/2}} \sum_k e^{ik\cdot x} \tilde{\phi}(k).
\end{equation}
The action can then be expressed in momentum space as
\begin{equation}\label{eq:lattice_scalar_fourier_action}
    S(\tilde{\phi}) = \frac{1}{2}\sum_k \left(k^2 + m^2\right) \left|\tilde{\phi}(k)\right|^2,
\end{equation}
where the lattice squared momentum is defined by
\begin{equation}
    k^2 = 4 \sum_\mu \sin^2 \frac{k_\mu}{2},\end{equation}
and sums over $k$ run over the finite-volume discrete momenta $k_\mu = 2\pi n_\mu/L$, $n_\mu \in \{0, 1, \dots, L-1\}$.

Because Eq.~\eqref{eq:lattice_scalar_fourier_action} is diagonal in $k$, the partition function reduces to a product over all Fourier modes $k$,
\begin{equation}
    Z = \prod_k \int d\tilde\phi(k) \, e^{-\frac12\left(k^2 + m^2\right) \left|\tilde{\phi}(k)\right|^2}.
\end{equation}
Since $\tilde{\phi}(-k) = \tilde{\phi}(k)^*$, the integration is to be taken over only the independent components, with each pair $\{k,-k\}$ contributing a single integral over $\operatorname{Re}\tilde{\phi}(k)$ and $\operatorname{Im}\tilde{\phi}(k)$. At the self-conjugate momenta $k \equiv -k \pmod{2\pi}$, $\tilde{\phi}(k)$ is real.

From this factorization, each independent mode is normally distributed:
\begin{equation}
    \tilde\phi(k) \sim \mathcal{CN}\left(0, \frac{1}{\omega(k)^2}\right), \quad \omega(k)^2 \equiv k^2 + m^2,
\end{equation}
where $\mathcal{CN}$ denotes a complex Gaussian distribution whose real and imaginary parts are independent with equal variance $[2 \omega(k)^2]^{-1}$.

Under a forward variance-preserving diffusion process with drift $u_\lambda(\phi_\lambda) = -\frac{1}{2}g_\lambda^2 \phi_\lambda$ (see Appendix~\ref{apx:euclidean_sde}), each mode evolves independently as
\begin{equation}
    \tilde{\phi}_\lambda(k) = \alpha_\lambda \tilde{\phi}_0(k) + \sigma_\lambda \eta_\lambda(k), \quad \eta_\lambda(k) \sim \mathcal{CN}(0, 1)
\end{equation}
with coefficients
\begin{equation}
    \alpha_\lambda = \exp\left[-\frac{1}{2}\int_0^\lambda g_s^2 ds\right], \quad \sigma_\lambda = \sqrt{1 - \alpha_\lambda^2}.
\end{equation}
The evolved configurations then also follow a Gaussian distribution for all diffusion times,
\begin{equation}\label{eq:phi_tilde_gaussian}
    \tilde{\phi}_\lambda(k) \sim \mathcal{CN}\left(0, \frac{\alpha_\lambda^2}{\omega(k)^2} + \sigma_\lambda^2\right).
\end{equation}
Eq.~\eqref{eq:phi_tilde_gaussian} implies a closed-form solution to the Fokker-Planck equation: up to normalization, the exact energy along the forward diffusion path is
\begin{equation}\label{eq:exact_free_field_energy}
    E_\lambda(\tilde{\phi}) = \frac{1}{2} \sum_k \frac{\omega(k)^2}{\alpha_\lambda^2 + \sigma_\lambda^2 \omega(k)^2} \left| \tilde{\phi}_\lambda(k)\right|^2
\end{equation}
which also provides analytical expressions for the force and Laplacian of the energy:\footnote{Here, $\nabla$ denotes the Wirtinger derivative with respect to $\tilde{\phi}(k)^*$. Equivalently, Eq.~\eqref{eq:free_field_score} gives the Fourier components of the real-space score, which is used in practice and obtained by an inverse Fourier transform.}
\begin{align}
    -\nabla E_\lambda(\tilde{\phi}) &= -\frac{\omega(k)^2}{\alpha_\lambda^2 + \sigma_\lambda^2 \omega(k)^2} \tilde{\phi}_\lambda(k), \label{eq:free_field_score} \\
    \nabla^2 E_\lambda(\tilde{\phi}) &= \sum_k \frac{\omega(k)^2}{\alpha_\lambda^2 + \sigma_\lambda^2 \omega(k)^2}. \label{eq:free_field_laplacian}
\end{align}

In our numerical applications, we consider 2D $L \times L$ lattices with $L \in \{16, 32, 64\}$. For each lattice volume, we follow a line of constant physics by fixing the bare mass to be $m = 2 / L$, 
giving approximate correlation lengths $\xi$ equal to 8, 16, and 32, respectively. The physical volume determined by the ratio $L/\xi \approx 2$ is therefore held fixed. The diffusion path is set by the It\^o SDE in Eq.~\eqref{eq:forward_ito_sde} with noise schedule
\begin{equation}
    g_\lambda = \sqrt{0.1 + (25 - 0.1)\lambda}
\end{equation}
so that $\alpha_\lambda \to 0$ and $\sigma_\lambda \to 1$ as $\lambda \to 1$. This leads to the standard normal distribution as $\lambda \to 1$, i.e., the action for the noise distribution is
\begin{equation}
    A(\phi) = \sum_x \phi(x)^2 / 2.
\end{equation}

For all models, we use the endpoint-constrained local-footprint architecture as described in Sec.~\ref{sec:ebm_parameterizations} with a $3 \times 3$ patch size. The interior energy density is computed on each site's local neighborhood using a neural network built from a circularly-padded $3 \times 3$ convolution. The 64 output channels of the convolution are concatenated with the embedded time $t_{\rm enc}$, followed by a linear projection to dimension 64. Finally, four residual blocks each consisting of three affine layers with SiLU activation functions are applied, followed by another SiLU activation and a linear projection from dimension 64 down to 1. The times are encoded using the procedure described above producing a vector $t_{\rm enc}$ with dimension 64.
The divergence for the flow is computed exactly from each patch's local Laplacian, which, when incorporated into Eq.~\eqref{eq:divergence_of_drift} for this diffusion process gives
\begin{equation}
\begin{aligned}
    \nabla \cdot \hat{v}_t(\phi_t) &= -\nabla \cdot \left[-\frac{1}{2}g_t^2\phi_t + \frac{1}{2}g_t^2\nabla \hat{E}_t(\phi_t)\right] \\
        &= \frac12 g_t^2 \left[L^d - \nabla^2 \hat{E}_t(\phi_t)\right].
\end{aligned}
\end{equation}

All free-field models are optimized by regressing directly onto the exactly known score given in Eq.~\eqref{eq:free_field_score}. In particular, the loss function to be minimized is the theoretical score-matching loss weighted by the noise level $\sigma_\lambda^2$ at each time,
\begin{equation}
\begin{aligned}
    \mathcal{L} &= \int_0^1 d\lambda \int \mathcal{D}\phi_\lambda \, \rho_\lambda(\phi_\lambda) \\
    &\qquad \qquad \qquad \times \sigma_\lambda^2\left\| \nabla \hat{E}_\lambda(\phi_\lambda) - \nabla E_\lambda(\phi_\lambda)\right\|_2^2.
\end{aligned}
\end{equation}
At each training step, the loss is estimated by sampling time steps $\lambda$ uniformly from the interval $[10^{-5}, 1]$ and free-field configurations $\phi_\lambda$ in batches of 1024 (512 for $L=64$). The free-field configurations are sampled exactly using the known Gaussian distribution in Eq.~\eqref{eq:phi_tilde_gaussian} along with the inverse Fourier transform in Eq.~\eqref{eq:inv_fourier_transform_discrete}. This zero-noise, data-free training approach is used to isolate the effect of the energy parameterization itself and to focus on the behavior of our proposed sampling strategies given an imperfect model. Each model is trained for a total of 160,000 training steps using the Adam optimizer~\cite{Kingma:2015} with base learning rate $10^{-4}$. The learning rate is decayed every 16 training steps following a cosine annealing schedule, reaching zero at the end of training. Training is carried out on a single NVIDIA GH200 GPU (the $L=64$ model on two GPUs) for a total duration of approximately 28 GPU-hours across all models.

\subsection{O(3) sigma model}\label{sec:o3_model_setup}
The action for the O(3) non-linear sigma model (NLSM) on the lattice is given by
\begin{equation}\label{eq:o3_nlsm_action}
    S(\phi) = -\beta \sum_x \sum_\mu \vec{\phi}(x) \cdot \vec{\phi}(x + \hat{\mu}),
\end{equation}
where $\beta$ is the inverse coupling constant and $\vec{\phi} \in \mathbb{R}^3$ with $\lVert \vec{\phi} \rVert = 1$. The theory features a global symmetry with respect to transformations $\vec{\phi}(x) \mapsto R \vec{\phi}(x)$, $R \in {\rm O}(3)$. As target parameters for this application, we consider $\beta \in \{ 1.4, 1.5, 1.6 \}$ with approximate infinite-volume correlation lengths $\xi$ respectively equal to 7, 11, and 19. We train and primarily evaluate the respective models on lattice geometries $L=16$, $L=32$, and $L=64$, corresponding approximately to a line of constant physics set by $L/\xi \approx 2$. Further transfers to larger volumes are also investigated.

Because the field variables $\vec{\phi}(x)$ live in a compact space, the diffusion process must be specified in a way that is compatible with this geometry. We use a standard Stratonovich formulation for the forward diffusion~\cite{Hsu:2002sam},
\begin{equation} \label{eq:langevin-sphere}
    d\vec{\phi}_\lambda = g_\lambda \left(\mathbb{I} - \vec{\phi}_\lambda \vec{\phi}_\lambda^\top\right) \circ  dW_\lambda,
\end{equation}
which leads to the Fokker-Planck equation
\begin{equation}
    \partial_\lambda E_\lambda - \langle \partial_\lambda E_\lambda \rangle = \tfrac{1}{2} g_\lambda^2 (\Delta E_\lambda - \lVert \nabla E_\lambda \rVert^2)
\end{equation}
with $\Delta$ the Laplace-Beltrami operator on the field configuration space (a product of spheres) and $\nabla E_\lambda$ the vector of tangent-space derivatives. Components of $\nabla E_\lambda$ are defined by
\begin{equation}
    \nabla^a_xf(\vec{\phi}) \equiv \frac{\partial}{\partial \omega}f\left(R_x(\vec{\phi}, \omega \, \vec{\phi} \times \hat{e}_a)\right)\Big|_{\omega=0},
\end{equation}
where $R_x(\vec{\phi},\vec{\theta})$ rotates $\vec{\phi}(x)$ by the rotation vector $\vec{\theta}$, i.e., by angle $\lVert \vec{\theta} \rVert$ about the axis $\hat{\theta}$.

To train via score matching it is also necessary to sample the heat kernel for arbitrary diffusion times $\lambda$. For each site $x$, the ${\rm O}(3)$ heat kernel at time $\lambda$ is given by
\begin{equation} \label{eq:heat-kernel-s2}
    K_\lambda(\vec{\phi}_\lambda | \vec{\phi}_0) \propto \sum_{\ell = 0}^{\infty} e^{-\ell(\ell + 1) \sigma_\lambda^2/2} P_\ell(\vec{\phi}_\lambda \cdot \vec{\phi}_0) (2 \ell + 1),
\end{equation}
where $P_\ell(\cdot)$ is the $\ell$-th Legendre polynomial.
For all numerical experiments, Eq.~\eqref{eq:heat-kernel-s2} is evaluated by truncating the sum at the 45\textsuperscript{th} term. The noise schedule is chosen to be
\begin{equation}
    g_\lambda = 2 \sqrt{3 \lambda^2} \quad \implies \quad \sigma_\lambda = 2 \lambda^{3/2},
\end{equation}
which results in a progressive introduction of noise over the interval $\lambda \in [0,1]$.

The majority of numerical results use a local-footprint EBM with footprint size $K = 5$ and a local energy network consisting of two residual blocks, with each residual block composed of two linear layers separated by SiLU activation functions, using hidden states of dimension $32$. Inputs to the local energy network are first pre-processed to produce all pairs of ${\rm O}(3)$ invariants $\vec{\phi}(x) \cdot \vec{\phi}(y)$ within each patch, except for a specific comparison to test quality of non-invariant models produced by passing raw vectors $\vec{\phi}(x)$ to the same local energy network instead. The time-encoding network uses an encoding vector $t_{\rm enc}$ of dimension $16$. Specific comparisons to other architectures are discussed below.

Training is performed with the composite denoising and target score-matching loss in Eq.~\eqref{eq:composite-loss} with ${\lambda_{\rm cut} = 0.1}$. For the O(3) model, the expression appearing in the target score identity can be expressed as
\begin{equation}
    \mathcal{T}_\lambda = -\phi_\lambda \times (\nabla_0 S(\phi_0) \times \phi_0).
\end{equation}
This expression can be derived as a small generalization of the Lie group case~\cite{debortoli:2024}, as shown in Appendix~\ref{app:o3-tsi}. In practice, training is carried out with the approximation $\mathcal{T}_\lambda \approx -\nabla_0 S(\phi_0)$.

Each model is trained for a total of 24 GPU-hours on a single H200 GPU, ranging between 144,000 and 500,000 training steps for the smallest to largest training volume targets. All training uses the Adam optimizer~\cite{Kingma:2015} with a fixed learning rate $3 \times 10^{-4}$.
At each training step, batches of size $512$ are sampled from an ensemble of $50,000$ configurations, which is refreshed every $10$ full passes through the data set.

For both HMC sampling and integrating the learned flow transformation, it is necessary to define an appropriate update along a vector field within the O(3) manifold. Throughout the numerical results in this work, we apply a standard geodesic rotation such that an update to an O(3) field configuration $\vec{\phi}$ along the vector field $\vec{v}$ with step size $dt$ is defined by
\begin{equation}
    \vec{\phi} \, \mapsto \, \cos(\alpha) \vec{\phi} + \sin(\alpha) \vec{v}/\lVert\vec{v}\rVert,
    \quad \alpha \equiv dt \lVert \vec{v} \rVert.
\end{equation}

\section{Results and Discussion}\label{sec:discussion}

\subsection{Model architecture and training results}

We first explore whether local-footprint models offer a wide enough receptive field to expressively represent energy functions over larger lattices and longer correlation lengths. To motivate the use of relatively small footprint sizes, we investigate the locality of the coupling between lattice sites under the free scalar field theory across diffusion time. The diffused energy for the free scalar field can be written in real space as
\begin{equation}
    E_\lambda(\phi) = \frac{1}{2}\sum_{x,y}\phi_\lambda(x) Q_\lambda(x - y)\phi_\lambda(y),
\end{equation}
where $Q_\lambda(x-y)$ is interpreted as the real-space coupling kernel between the fields at different lattice sites, and can be obtained by applying an inverse Fourier transform to the coefficient of $|\tilde{\phi}_\lambda(k)|^2$ in Eq.~\eqref{eq:exact_free_field_energy}. 
Rewriting the frequency as
\begin{equation}
    \omega(k)^2 = \frac{1}{\sigma_\lambda^2}\left(\alpha_\lambda^2 + \sigma_\lambda^2 \omega(k)^2\right) - \frac{\alpha_\lambda^2}{\sigma_\lambda^2}
\end{equation}
gives the partial fraction decomposition of $\tilde{Q}_\lambda$ into
\begin{equation}\label{eq:partial_fraction_kernel}
    \tilde{Q}_\lambda(k) = \frac{1}{\sigma_\lambda^2} - \frac{\alpha_\lambda^2}{\sigma_\lambda^4}\frac{1}{k^2 + M_\lambda^2},
\end{equation}
where we define an effective mass $M_\lambda$ by
\begin{equation}\label{eq:effective_mass}
    M_\lambda^2 = m^2 + \alpha_\lambda^2/\sigma_\lambda^2.
\end{equation}
The inverse Fourier transform of Eq.~\eqref{eq:partial_fraction_kernel} then yields
\begin{equation}\label{eq:analytical_continuum_kernel}
    Q_\lambda(r) = \frac{1}{\sigma_\lambda^2}\delta(r) - \frac{\alpha_\lambda^2}{\sigma_\lambda^4}G_{M_\lambda}(r), \quad r \equiv x - y,
\end{equation}
where $G_{M_\lambda}$ is the free massive propagator. The effective mass $M_\lambda$ sets the length scale of $Q_\lambda$, whose tails fall off as $e^{-M_\lambda r}$. Near the target, $\sigma_\lambda \to 0$, so ${M_\lambda \to \infty}$ and the range shrinks to zero. Near the prior, ${\alpha_\lambda \to 0}$, so ${M_\lambda \to m}$ and the range grows to the correlation length of the target theory; however, the coefficient of the propagator in Eq.~\eqref{eq:analytical_continuum_kernel} also vanishes in this limit, thereby attenuating the long tails.

\begin{figure}
    \centering
    \includegraphics[width=0.95\linewidth]{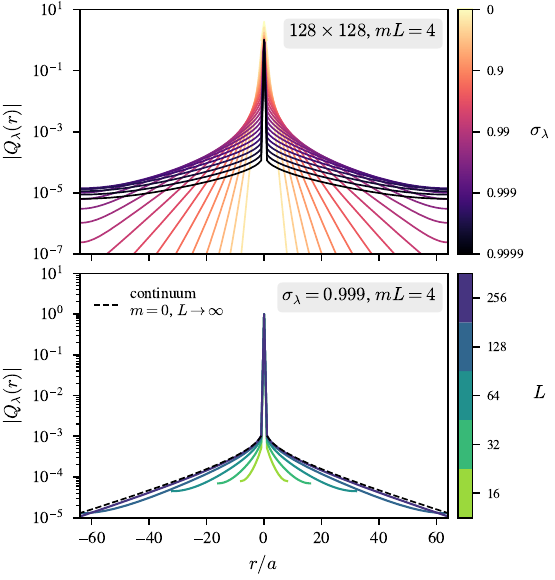}
    \caption{Real-space coupling kernel $Q_\lambda$ of the exact diffused energy $E_\lambda$ for a free scalar field plotted against the spacetime separation $r \equiv x - y$. The top panel shows $Q_\lambda$ for a $128 \times 128$ lattice over different noise levels $\sigma_\lambda$. The exact energy is highly localized for all diffusion times, with a non-local tail 3-5 orders of magnitude smaller than the peak for intermediate diffusion times. The bottom panel shows $Q_\lambda$ for a fixed $\sigma_\lambda = 0.999$ and $mL=4$ over different lattice volumes approaching the continuum limit, with the dashed black curve showing the continuum result in Eq.~\eqref{eq:analytical_continuum_kernel}.}
    \label{fig:kernel_localization}
\end{figure}

To further investigate the locality structure, we compute $Q_\lambda$ on the lattice numerically and plot its spread over distance on the lattice in Fig.~\ref{fig:kernel_localization}.
At every diffusion time, we find that the exact energy is highly local, indicated by the sharp peak in $|Q_\lambda(r)|$ near ${r=0}$. At $\lambda=0$, $\sigma_\lambda = 0$ and $Q_0$ is the massive nearest-neighbor Laplacian, while at $\lambda=1$, $\sigma_\lambda \to 1$ and $Q_1(r) = \delta(r)$. The only non-locality in $Q_\lambda$ arises for intermediate diffusion times where ${\sigma_\lambda \approx 0.99\text{--}0.999}$; in this range $Q_\lambda$ has the heaviest tails, reaching about $10^{-5}$ across the finite lattice domain shown. Additionally, varying the correlation length by fixing $mL=4$ and varying lattice sizes $L \in \{16, 32, 64, 128, 256\}$ has minimal effect on the range of $Q_\lambda$. The exact continuum limit curve derived in Eq.~\eqref{eq:analytical_continuum_kernel} is also shown for reference at $\sigma_\lambda = 0.999$, demonstrating that the coupling is in fact exponentially localized in this limit, as suggested by the massive propagator appearing in Eq.~\eqref{eq:analytical_continuum_kernel}.

\begin{figure}
    \centering
    \includegraphics[width=0.9\linewidth]{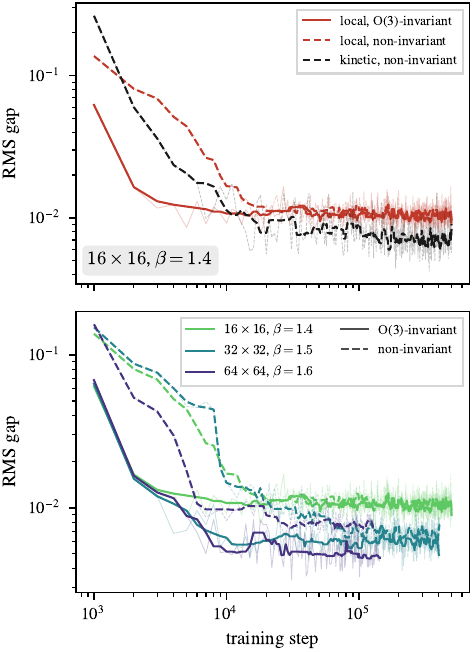}
    \caption{RMS gap between forward and reverse energy densities during training for the ${\rm O}(3)$ model. Top: local-footprint models compared with a kinetic EBM built from a non-invariant U-Net at $16 \times 16$, $\beta=1.4$. Bottom: the local-footprint model with (solid lines) and without (dashed lines) built-in ${\rm O}(3)$-invariance at $(L,\beta) = (16,1.4), (32, 1.5), (64, 1.6)$.}
    \label{fig:o3_model_comparison}
\end{figure}

A related question is whether the local-footprint structure can in practice be trained to a high-quality approximation of the diffusion energy, when compared to a black-box, long-range parameterization. The free scalar theory arguments above already demonstrate that a local footprint can give exponential accuracy for this theory. To check whether this statement translates to an interacting theory, we compare training a local-footprint model versus a kinetic energy parameterization which uses a U-Net vector field for the ${\rm O}(3)$ NLSM. The latter in principle can encode long-range correlations through convolutions applied in the coarse levels within the U-Net construction.
This comparison is shown in the top panel of Fig.~\ref{fig:o3_model_comparison}, where the U-Net kinetic energy model is compared to both the invariant and non-invariant local-footprint models. Shown in the figure as a measure of quality during training is the root-mean-squared (RMS) gap between the measured energy density of sampled configurations $\phi_{\lambda}$ generated from the true forward diffusion process versus those generated by sampling the learned energy function $\hat{E}_\lambda$ at various points during training; the floor in this quantity is set by fluctuations in the finite batches used to estimate the energy density, with this batch size chosen so that early training dynamics are clearly revealed. Both variants of local footprint models converge to a similar RMS gap, and also reach the optimum value more quickly than the U-Net kinetic energy model. While the U-Net kinetic energy model achieves a better optimal value, this comes at the cost of being unable to efficiently estimate its Laplacian $\nabla^2 \hat{E}_t$.

We finally explore whether it is beneficial to incorporate the exact ${\rm O}(3)$ symmetry in the ${\rm O}(3)$ NLSM models.
We compare two classes of local-footprint models, which differ only in the preprocessing network used to collect degrees of freedom in each patch into the local hidden vector $\vec{\phi}_{\rm enc}$ of dimension $n_h$. The first, non-invariant option uses a standard convolution which takes $n_h$ linear combinations of the field variables within the patch; the second, invariant option first constructs all pairwise inner products between field variables within the patch, then takes $n_h$ linear combinations of these invariants. The bottom panel of Fig.~\ref{fig:o3_model_comparison} compares the training dynamics of these two networks. The figure reveals that the exactly symmetric models converge more quickly and converge to a better overall model quality for the larger values of $\beta$. Together, these results motivate our focus on only the (${\rm O}(3)$-invariant) local-footprint models for all further results in the scalar field theory and ${\rm O}(3)$ NLSM below.

\subsection{Sampling the learned path using SNFs}
We first consider sampling the trained diffusion models using the SNF approach with various numbers of annealing steps, $n$. In all cases, each sample is evolved independently by interleaving HMC updates with flow-transport steps, as outlined in Algorithm~\ref{fig:main-algo}. We denote the total number of independent samples generated for any given target by $N$.
For sampling, we use a uniform grid of $n$ annealing steps in $t$. At each step, the $N$ samples first undergo $n_{\rm MCMC}$ HMC updates targeting the learned energy $\hat{E}_t$ at the current time, after which they are transported to the next timestep by a single RK4 step of the probability-flow ODE in Eq.~\eqref{eq:flow-dt}. The log-Jacobian is accumulated by the same RK4 step using the exact divergence of the local-footprint model, and the log-weights are updated according to Eq.~\eqref{eq:snf-weights}. For the free scalar field theory, we sample $N=256$ configurations with $n_{\rm MCMC}=10$. Our HMC trajectories have length 8.0 for $L=16$ and $16.0$ for $L=32, 64$ with 20, 64, and 80 leapfrog steps respectively. For the ${\rm O}(3)$ NLSM, we sample $N=256$ ($N=1024$ for $\beta=1.4$, $n=100$) configurations with $n_{\rm MCMC}=1$. In this case, HMC trajectories have length $0.5$ with $10$ leapfrog steps.

As a comparison, we also perform sampling
for the same targets without the flow (i.e., using AIS sampling) for both theories. In the case of scalar field theory, the settings are the same as SNF sampling. For the ${\rm O}(3)$ NLSM, we sample $N=1024$ configurations for a higher-statistics comparison in all cases, while using the same HMC settings.

For sequential annealing methods, the effective sample size (ESS) provides a direct measure of sample quality. In terms of the sampled weights, it is defined by
\begin{equation}\label{eq:ess}
    \text{ESS} = \frac{\left(\frac{1}{N}\sum_i w_i\right)^2}{\frac{1}{N}\sum_i w_i^2} \in[0,1].
\end{equation}
In general, the ESS represents the fraction of the sampled ensemble of size $N$ that makes non-trivial contributions to final expectation values; a perfect ESS of 1 is obtained when all weights are equal.
The ESS can also be related to the variance of the log-weights $\log{w}$ themselves: for roughly log-normal $w$, $\text{ESS} \approx \exp(-\Var[\log w])$. While not directly a measure of sampling efficiency, the variance $\Var[\log w]$ typically has lower noise than measurements of the ESS and can be measured even when the ESS is very small. This variance also directly relates to the theoretical quantities laid out in earlier sections.

\begin{figure}
    \centering
    \includegraphics[width=0.9\linewidth]{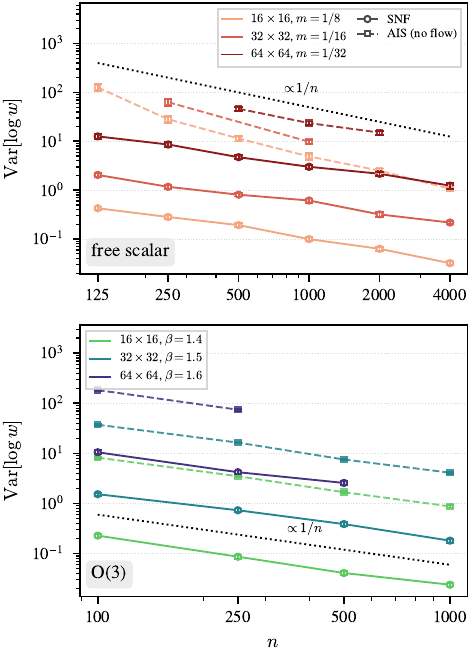}
    \caption{Variance of the log weights plotted against the number of annealing steps for SNF (solid lines) and AIS (dashed lines) on $16 \times 16$, $32 \times 32$, and $64 \times 64$ lattices, for the free theory with $m=2/L$ (top) and the ${\rm O}(3)$ model with $\beta=1.4,1.5,1.6$ (bottom).
The same learned energies and HMC settings are used for both AIS and SNF, with one exception:
    HMC trajectories for the free theory on the $64 \times 64$ lattice are integrated with twice the leapfrog steps for AIS, without which AIS walkers are consistently rejected near the target.
The 1--2 orders of magnitude improvement between the AIS and SNF results comes from using the nearly optimal flow derived from the score.}

    \label{fig:snf_vs_ais}
\end{figure}

In Fig.~\ref{fig:snf_vs_ais}, we compare the scaling of $\Var[\log w]$ for SNF versus AIS in the free scalar field theory and the ${\rm O}$(3) model. 
Incorporating the learned flow reduces the reweighting variance by one to two orders of magnitude relative to AIS over the range of $n$ studied.
The residual variance under SNF comes from model error in $\hat{E}_t$.

\begin{figure}
    \centering
    \includegraphics[width=0.9\linewidth]{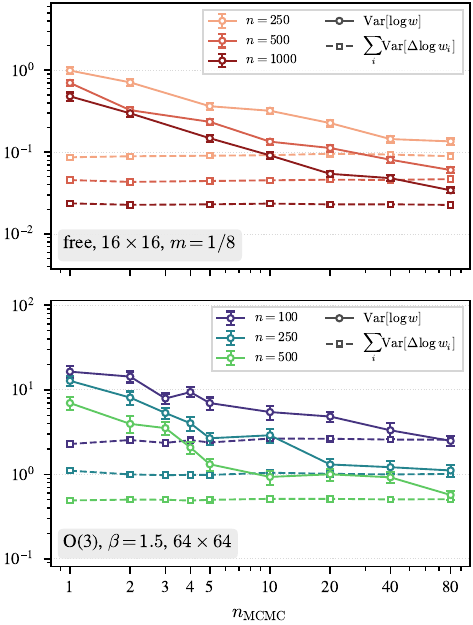}
    \caption{Total variance of the log weights $\Var[\log w]$ (solid lines) compared with the sum of per-step increment variances $\sum_i \Var[\Delta \log w_i]$ (dashed line) against the number of HMC updates per step $n_{\rm MCMC}$ for different numbers of annealing steps $n$ for the free scalar field theory (top) and the ${\rm O}(3)$ model (bottom). The per-step sum scales as $1/n$; the excess of the total variance reflects correlations between steps, which vanish as $n_{\rm MCMC}$ increases.}
    \label{fig:snf_var_vs_n_mcmc}
\end{figure}

In both theories, $\Var[\log w]$ decreases monotonically with $n$ and the ESS rises correspondingly (see App.~\ref{app:scaling}), enabling systematic refinement of sample quality post-training. For SNF sampling from the scalar field theory, we observe a slower scaling than the asymptotic $1/n$ expected in the quasi-static limit. This is explained by correlations between steps: writing ${\log w = \sum_i \Delta \log w_i}$, the variance splits into per-step contributions and the covariance between steps as
\begin{equation}\label{eq:var_covar_logw}
\begin{aligned}
    \Var[\log w] &= \sum_i \Var[\Delta \log w_i] \\ &\quad + 2 \sum_{i<j}\operatorname{Cov}[\Delta \log w_i, \Delta \log w_j].
\end{aligned}
\end{equation}
In Fig.~\ref{fig:snf_var_vs_n_mcmc}, we compare the total variance with the per-step sum for the two theories. The per-step sum is independent of $n_{\rm MCMC}$ and follows an ideal $1/n$ scaling, indicating that the excess in the full variance arises because the HMC updates per SNF step do not fully decorrelate successive log-weight increments. Increasing the number of HMC updates reduces the variance systematically towards the floor set by the independent sum. For the ${\rm O}(3)$ model, the ratio of the total to the per-step sum decreases from 7--14 at $n_{\rm MCMC}=1$ to 1.0--1.1 at $n_{\rm MCMC}=80$ ($n=100$--500); for the free theory, it decreases from 11--20 at $n_{\rm MCMC}=1$ to 1.3--1.5 at $n_{\rm MCMC}=80$ ($n=250$--1000).

The residual correlations also affect SNF sampling more strongly than AIS (no-flow) sampling, as observed for example by the nearly perfect $1/n$ scaling for AIS.
We propose the following qualitative explanation. Including the flow transformation to move from AIS to SNF sampling already significantly reduces the variance of each log-weight increment. Because the AIS variance is expected to be dominated by UV fluctuations, we also expect these to be the most improved by the inclusion of the flow transformation; this leaves the more slowly mixing intermediate and IR modes as a larger contribution to the residual variance and to the covariance term in Eq.~\eqref{eq:var_covar_logw}.
Although the quasi-static limit is approached less rapidly once the flow is applied, SNF sampling nonetheless consistently improves over AIS by up to two orders of magnitude in $\Var[\log w]$ across all $n$ studied.

\subsection{Sampling the learned path using T-REX}

For each lattice size, we use a uniform ladder of $n+1$ chains, targeting the learned distribution for times ${t_k = k/n}$, with $k=0, \dots, n$.
One sweep consists of $n_{\rm MCMC}$ HMC trajectories applied on each chain, targeting the respective learned energy $\hat{E}_{t_k}$, followed by one round of swap proposals between neighboring chains. Swaps alternate between even pairs $(t_{2j}, t_{2j+1})$ and odd pairs $(t_{2j+1}, t_{2j+2})$ on successive sweeps, so each pair is proposed on every other sweep. In T-REX, each proposed configuration is transported to its new chain by a single RK4 step of the probability-flow ODE in Eq.~\eqref{eq:flow-dt}; the log-Jacobian required for the acceptance probability is also computed by a single RK4 step. REX uses the same schedule without the transport step.

For scalar field theory results, the following settings are applied.
The ladder is initialized by drawing independent samples from the prior at $t=0$ and integrating the learned probability-flow ODE to each intermediate time $t_k$ with 100 RK4 steps, chosen so that the integrator converges well for the largest value, $t = 1$. Each sweep uses $n_{\rm MCMC} = 2$ HMC trajectories between swap proposals.
The swap variance profiles are measured over 2000 sweeps after 500 warmup sweeps, using HMC trajectories of length $4.0$ with 8, 10, and 18 leapfrog steps for $L=16, 32,64$, giving acceptance rates of 0.57, 0.50, and 0.67, respectively.

For the ${\rm O}(3)$ NLSM, the following settings are applied.
The ladder is initialized by drawing independent samples from the prior (uniform noise) for all chains. The majority of REX and T-REX runs are evaluated using 1024 sweeps, each of which proposes one (flowed) swap update following one HMC trajectory of length $0.5$, integrating using $10$ leapfrog steps. The only exceptions to these settings are for $L \geq 96$, for which HMC trajectories are evolved for length $0.4$ using $20$ leapfrog steps, and specifically for $L=128$ results which use fewer overall sweeps, respectively 512 sweeps for the $n=50$ case and 256 sweeps for the $n=100$ case.

For measurements of observables and metrics, 100 thermalization steps are skipped before evaluation in all cases except for the $\beta=1.6$, $L=64$ ${\rm O}(3)$ model, for which 824 steps are skipped for observables, leaving 200 samples for measurements.
Because successive measurements remain correlated in both replica-exchange methods, we use a blocked bootstrap method to estimate uncertainties.
For the scalar field theory, errors on the swap-variance profiles are estimated with 200 resampling steps using blocks of 50 sweeps, while errors on observables are estimated with 1000 resampling steps using blocks of 100 sweeps.
For the ${\rm O}(3)$ model measurements, errors are by default estimated with 1000 resampling steps using blocks of 50 sweeps. Two exceptions are made: $\beta=1.6$, $L=64$ uses blocks of 24 sweeps, while $\beta=1.6$, $L=128$, uses blocks of 18 sweeps, in both cases to ensure sufficiently many blocks are available for bootstrap estimates.

\begin{figure}
    \centering
    \includegraphics[width=0.9\linewidth]{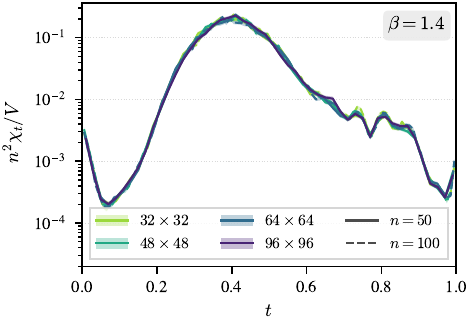}
    \caption{Rescaled variance of the swap log-weights versus chain-pair midpoints $t$ for the ${\rm O}(3)$ model at fixed coupling $\beta=1.4$ over four lattice volumes with $n=50$ and $n=100$. All curves collapse onto one another when rescaled, indicating that the variance grows proportionally to the volume $V$.}
    \label{fig:logw_swap_fixed_beta_o3}
\end{figure}

As introduced in Eq.~\eqref{eq:small_dt_expansion_var_logw}, $\chi_t$ is defined as one half the variance of the swap log-weights at time $t$. It relates closely to sequential annealing metrics for small $\Delta t$ according to
${\chi_t \approx \Delta t^2 \Var[\xi_t]}$. Since $\xi_t$ is a sum of local contributions over lattice sites, $\Var[\xi_t] \propto V$ and the swap variance is expected to scale as $V/n^2$.
To test this theoretical scaling, we show in Fig.~\ref{fig:logw_swap_fixed_beta_o3} the rescaled quantity $n^2 \chi_t / V$ over the T-REX ladder for the ${\rm O}(3)$ model at $\beta = 1.4$ across lattice sizes $L \in \{32, 48, 64, 96\}$. All of these volumes are evaluated by transferring the model trained at $L=16$, without retraining. The rescaled quantities collapse to the same curve across $n$ and $L$, confirming the expected scaling. This scaling implies that the number of chains $n$ should scale with $\sqrt{V}$ to obtain fixed swap acceptance using T-REX, as compared with $n \propto V$ in SNF.
However, this comparison must also take into account the fact that T-REX chains are autocorrelated. We provide a more detailed cost comparison in the following section.

\begin{figure}
    \centering
    \includegraphics[width=0.9\linewidth]{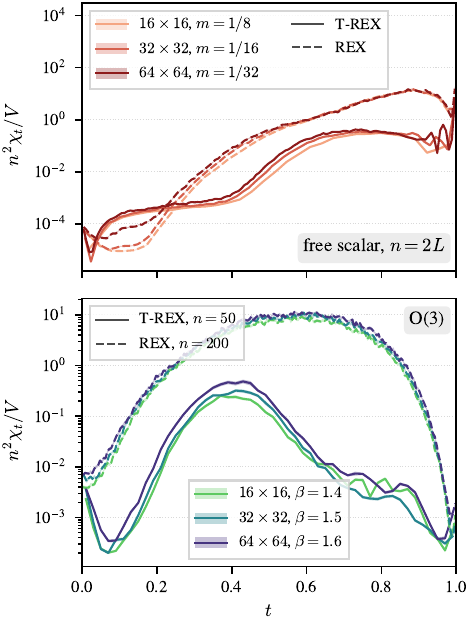}
    \caption{The rescaled variance of the swap log-weights versus the midpoint $t$ of each pair of neighboring chains for (top) the free scalar theory and (bottom) the ${\rm O}(3)$ model. Solid lines are T-REX and dashed are REX. Results are shown across different lattice sizes with approximately fixed physical volume.
Shaded bands show block-bootstrap errors.}
    \label{fig:logw_swap_fixed_physical_vol}
\end{figure}

A more physically meaningful comparison is following the continuum limit along a line of constant physics, as studied also in the case of SNF sampling. The rescaled local variance contributions, $n^2 \chi_t/V$, are shown for both scalar field theory and the ${\rm O}(3)$ model in Fig.~\ref{fig:logw_swap_fixed_physical_vol}. We first find that T-REX reduces the swap variance by one to three orders of magnitude over most of the replicas, relative to (unflowed) REX. For scalar field theory, improvements are particularly concentrated in the range $t \gtrsim 0.25$, while for the ${\rm O}(3)$ model improvements are observed over the entire range of annealing time. The learned drift thus significantly improves the overlap between neighboring chains, so that T-REX needs a factor of about 4--5 times fewer chains than REX in the free theory, and about 7--9 times fewer chains in the ${\rm O}(3)$ model, to achieve a given average swap acceptance rate. Below $t \approx 0.23$ in the free scalar theory, there is a crossover where the variance under REX falls below that of T-REX at every $L$; however, both variances are negligible in this region, as both methods yield a swap acceptance rate of 99\%. We attribute this behavior to model imperfection: since $\hat{E}_t$ is trained globally with a single loss across $t$, there is no guarantee that the minimizer of the loss for a limited model ansatz will improve the swap acceptance rate over the entire range of $t$.

The same variance profiles also show that the T-REX swap acceptance rate is worst in the middle region ($0.2 \lesssim t \lesssim 0.6$) for the ${\rm O}(3)$ model, whereas for the free theory it is worst near the target ($t \gtrsim 0.6$). We do not attribute significant physical meaning to these effects: the most substantial transport costs are determined both by how quickly the distribution evolves under the artificially introduced diffusion equation, including the arbitrary choice of $g_\lambda$, and by how well our models reproduce the intermediate probability densities. However, the fact that these costs are not uniformly distributed indicates that a constant-factor improvement could be gained by rescheduling the annealing time $t$; we show this effect for the $\phi^4$ theory in Sec.~\ref{sec:phi4_trex_demo}.

While T-REX provides a large factor gain over REX in practice, the scaling of the cost is also important. We see that the curve collapse of the rescaled swap log-weight variance is not perfect, as there is a marginally higher variance than naive degrees-of-freedom scaling would suggest across the annealing time range for both theories. For the case of the free scalar field theory, this largely affects the parts of the diffusion path that make smaller contributions to the total refresh time, while for the ${\rm O(3)}$ NLSM, the growth with decreasing lattice spacing manifests in the most challenging part of the annealing time range.
We quantify the resulting costs next.

\subsection{Lattice spacing effects and cost}\label{sec:cost_comparison}

\begin{figure}
    \centering
    \includegraphics[width=0.9\linewidth]{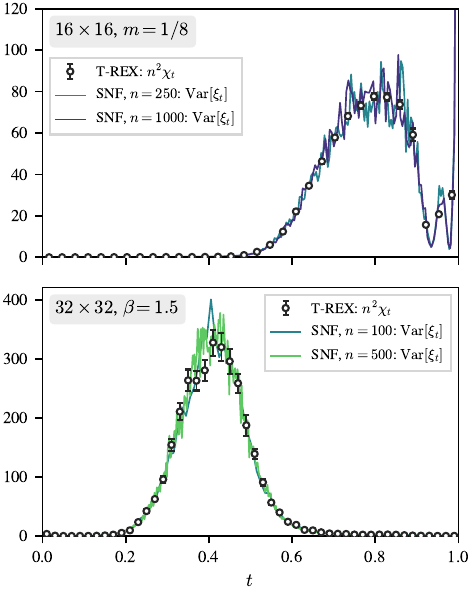}
    \caption{Two independent estimates (SNF: $\Var[\xi_t]$, T-REX: $n^2\chi_t$) for the variance of the instantaneous log weight increment $\xi_t$ along annealing time $t$ in the free theory on a $16 \times 16$ lattice (top) and the ${\rm O}(3)$ model at $\beta=1.5$ on a $32 \times 32$ lattice (bottom). Results obtained from equilibrium T-REX samples are compared to SNF, demonstrating that the latter is nearly in equilibrium and that results translate well between the two sampling methods.}
    \label{fig:var_xi_t}
\end{figure}
In both theories, $\Var[\xi_t]$ measured from SNF sampling aligns well with $n^2 \chi_t$ measured from T-REX, as depicted in Fig.~\ref{fig:var_xi_t}. This indicates both that the SNF samples remain close to equilibrium at every $t$ and that $n$ is large enough in both cases to suppress high-order corrections. More importantly, this demonstrates that both samplers are governed by the same thermodynamic metric along the path, such that results on scaling, annealing-time schedules, and projected variance/acceptance rates can be easily connected from one method to the other.

To compare the overall sampling cost in either case, we first note that the cost of T-REX is set both by the number of chains $n+1$, which sets the cost of each sweep, and the refresh time $\tau$, which sets the number of sweeps between independent configurations arriving at the target chain. At fixed acceptance, Figs.~\ref{fig:logw_swap_fixed_beta_o3} and \ref{fig:logw_swap_fixed_physical_vol} imply that one must set $n \propto \tilde{\Lambda}$. Combining this with the asymptotic scaling $\tau \sim \tilde{\Lambda}$, the expected total number of updates required to acquire independent samples from T-REX scales as $n \tau \sim \tilde{\Lambda}^2$. 
 In the case of SNF sampling, the same quantity sets the lower bound on the variance by Eq.~\eqref{eq:optimal_annealed_action}, $\Var[\log w] = \tilde{\mathcal{S}}/n \geq \tilde{\Lambda}^2/n$.
Maintaining a fixed $\Var[\log w]$ therefore requires $n \propto \tilde{\Lambda}^2$ annealing steps per independent sample, matching the scaling for T-REX.
In all cases, naive degrees-of-freedom scaling leads to $\tilde{\Lambda}^2 \sim V$ so that the number of steps required per independent sample scales linearly with the lattice-units volume when all else is held constant.

Both methods also benefit from applying the same optimal annealing schedule, which places the annealing times at equal increments of the path length $\tilde{\Lambda}$; however, the benefits for each method manifest in different ways. For SNF, it reduces the variance at fixed $n$ from $\tilde{\mathcal{S}}/n$ to its lower bound $\tilde{\Lambda}^2/n$, which is reached when the path is traversed at constant speed. For T-REX, the asymptotic refresh time $1 + \tilde{\Lambda}/\sqrt{\pi}$ is independent of the schedule, since the length $\tilde{\Lambda}$ is a parameterization-independent geometric property of the curve $t \mapsto \rho_t$. Instead, the optimal schedule reduces the number of chains needed to approach this limit, i.e., it sets the threshold $n$ where the autocorrelation time is nearly saturated. This is the effect seen for the $\phi^4$ theory in Sec.~\ref{sec:phi4_trex_demo}, where the optimal schedule reduces the refresh time at the same $n$.

\begin{figure}
    \centering
    \includegraphics[width=0.9\linewidth]{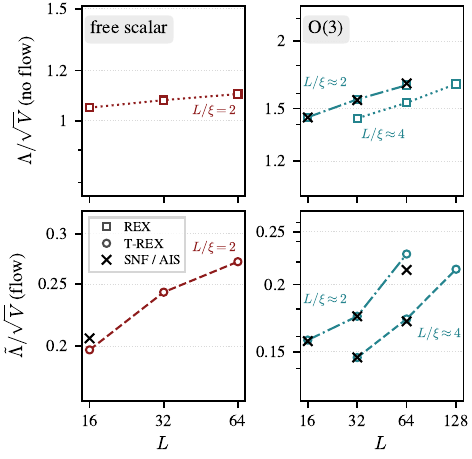}
    \caption{Rescaled thermodynamic path length for the unflowed ($\Lambda/\sqrt{V}$) and flowed ($\tilde{\Lambda}/\sqrt{V}$) annealing paths in the free scalar theory (left) and the ${\rm O}(3)$ model (right). Points are estimated from REX, T-REX, and SNF data, as discussed in the main text.
Connected sets of points represent flowed and unflowed paths at approximately fixed physical volume, defined by fixing the ratio $L/\xi$ between the lattice size and correlation length,
for which the continuum limit is approached by increasing the lattice size $L \to \infty$.
    Each pair of vertical panels spans the same width in decades, so slopes within each theory are directly comparable.}
    \label{fig:path_lengths}
\end{figure}

In terms of total cost, each MCMC or flow update acts on a lattice of volume $V$, so the computational cost per update also features a basic scaling proportional to $V$. Notably, the model did not need to be extended for any of the studies here---a fixed $3 \times 3$ or $5 \times 5$ footprint was used for the scalar and ${\rm O}(3)$ theories, respectively---and this cost was thus constant.

To understand the path length more precisely from numerical data, we note that the path length can be respectively estimated from SNF or T-REX according to~\cite{Syed:2024}
\begin{equation} \label{eq:path-length-estimators}
    \tilde{\Lambda} = \lim_{\dt \to 0} \sum_{k=1}^n \sqrt{\Var[\Delta \log w_{t_k}]} = \lim_{\dt \to 0} \sum_{k=1}^{n} \sqrt{\chi_{t_k}}.
\end{equation}
Analogous estimators for the unflowed path length $\Lambda$ are given by evaluating the equation above using AIS or REX.
In the simplest case of fixing the physical coupling while scaling the physical volume, shown for example in Fig.~\ref{fig:logw_swap_fixed_beta_o3}, we observe the expected trivial scaling $\tilde{\Lambda} \sim \sqrt{V}$. This implies an overall cost scaling of $\tilde{\Lambda}^2 V = O(V^2)$, as has been observed in previous works on SNF and non-equilibrium sampling~\cite{Bulgarelli:2024brv}.

More interesting is the scaling along a line of constant physics towards the continuum limit.
We define such a line by holding the ratio of lattice size $L$ and correlation length $\xi$ approximately fixed, as above.
Approaching the continuum, the measured $\tilde{\Lambda}$ grows slightly faster than $\sqrt{V}$, as indicated for example by the imperfect collapse of the swap variances seen above in Fig.~\ref{fig:logw_swap_fixed_physical_vol}.
We quantify the deviation from the naive scaling estimate for the total thermodynamic path length by considering its value divided by $\sqrt{V}$.
The results for both theories are shown in Fig.~\ref{fig:path_lengths}. In addition to the previously studied value $L/\xi \approx 2$ for the scalar field theory and ${\rm O}(3)$ model, simulations at twice the lattice size are added for the ${\rm O}(3)$ model, using volume transfer to evaluate trained models directly at the larger volume. The latter dataset uses lattice sizes $L = 32, 64, 128$ at the same $\beta \in \{1.4, 1.5, 1.6\}$, corresponding to $L/\xi \approx 4$. Across all measurements, we find that the ratio grows weakly towards the continuum according to the anomalous scaling $\tilde{\Lambda} \sim V^{p+1/2}$,  with a power $p \lesssim 0.15$. The cost per independent sample along these lines of constant physics therefore grows marginally faster than the naive $V^2$ scaling observed at fixed coupling.

\subsection{Scalar \texorpdfstring{$\phi^4$}{phi4} demo}\label{sec:phi4_trex_demo}
As an additional demonstration of our method, we apply T-REX sampling in the 2D $\phi^4$ theory, using the action in Eq.~\eqref{eq:lattice_scalar_action} with parameters in the broken phase.
The $\phi^4$ target we investigate uses action parameters $m^2 = -4.21$, $\lambda=1.11$ chosen by a parameter scan to identify values where HMC tunnels freely between the two sectors on a $16 \times 16$ volume, while remaining frozen on a $48 \times 48$ volume. With this setup, we use efficient HMC samples on the $16 \times 16$ volume for training, then evaluate the efficiency of the annealed diffusion sampling method for the challenging $48 \times 48$ target volume. Our diffusion settings are the same as for the free theory, i.e., a variance-preserving diffusion process with noise schedule $g_\lambda = \sqrt{0.1 + 24.9\lambda}$.

We use the same local-footprint model architecture as for the free scalar field theory.
The model is trained on an ensemble of 16{,}384 configurations, generated using 256 independent HMC chains with 64 configurations each; saved configurations are spaced 400 trajectories apart, after skipping 2000 thermalization trajectories per chain. The last 1024 configurations are held out to monitor a validation loss, leaving 15{,}360 for training. Each HMC trajectory has length $2.0$, integrated using 11 leapfrog steps, yielding an average acceptance rate of 0.66. The training set experienced 2673 tunneling events in total,
giving a sufficiently representative ensemble for the target theory.
Using this setup, the model is trained for 10,000 passes through the full dataset in 15 batches of 1024 configurations, giving a total of 150,000 gradient updates. To reduce overfitting to the finite training ensemble, each batch is refreshed by one HMC update before each training step, using a trajectory of length 0.5 integrated with 5 leapfrog steps. The refreshed configurations are not stored. The model is trained according to the weighted denoising score matching objective, i.e.~Eq.~\eqref{eq:composite-loss} with $\lambda_{\rm cut} = 0$. To avoid numerical instability near $\lambda = 0$, the diffusion time is sampled over the restricted range $\lambda \in [10^{-5}, 1]$. The remaining settings are the same as for the free field example.

To further increase our learned model quality, we manually symmetrize the energy at evaluation time,
\begin{equation}\label{eq:symmetrize_model}
    \hat{E}_t(\phi) \rightarrow \frac{1}{2}\left( \hat{E}_t(\phi) + \hat{E}_t(-\phi)\right).
\end{equation}
We note that this is not equivalent to simply symmetrizing our data by random $\mathbb{Z}_2$ flips, which would by brute force solve tunneling challenges in all cases, including for HMC. Instead, a symmetrized energy function leads only to an incrementally better flow transport and swap acceptance rate for T-REX sampling, so that it provides a non-trivial check of the efficiency of this approach.
\begin{figure}
    \centering
    \includegraphics[width=0.9\linewidth]{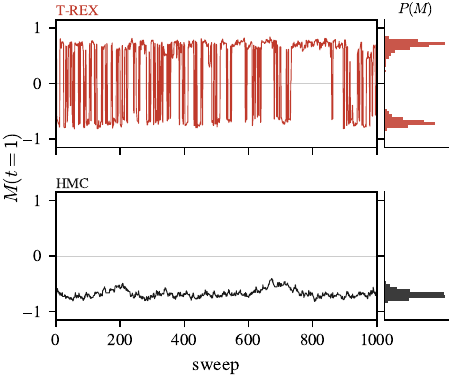}
    \caption{Magnetization $M$ of the target configuration ($t=1$) against sweeps for the broken-phase $\phi^4$ target on a $48 \times 48$ lattice. T-REX sampling using the symmetrized learned energy is compared to an HMC chain directly sampling the target action. Traces show the first 1000 sweeps; histograms show the distribution of $M$ values over all 2000 sweeps used in each case. HMC remains in one mode, whereas T-REX tunnels repeatedly and reproduces the two magnetization peaks of the broken phase.
}
    \label{fig:phi4_magnetizations}
\end{figure}

We first sample the $48 \times 48$ target using T-REX with a uniform schedule, using a ladder of $n+1=97$ chains at annealing times $t_k = k/n$, where $k \in \{0, \dots, n\}$. Each chain targets the distribution for the respective symmetrized energy $\hat{E}_t$ given in Eq.~\eqref{eq:symmetrize_model}. The ladder is initialized by drawing independent samples from the prior and integrating the learned probability flow ODE to each $t_k$ with 100 RK4 steps. One T-REX sweep consists of two HMC trajectories per chain (trajectory length 2.0 integrated with 18 leapfrog steps), followed by one round of swap proposals that alternate between even and odd pairs of neighboring chains as in Sec.~\ref{sec:annealed_sampling_methods}. In total, we run 2000 sweeps after 500 warmup sweeps.
For comparison, we run 97 independent HMC chains on the exact target action, one per T-REX chain so that the cost per sweep matches, using the same trajectory settings. Fig.~\ref{fig:phi4_magnetizations} and the $\langle M\rangle$ in Table~\ref{tab:phi4_trex_obs} show a single one of these chains, while the observables in App.~\ref{app:phi4_extra_obs} average over all 97.

We characterize the sampled configurations by their magnetization $M = \sum_x \phi(x)/L^2$. The distribution of $M$ for our parameters is bimodal with peaks around $\pm 0.56$. The $\mathbb{Z}_2$ symmetry of the theory guarantees $\langle M\rangle = 0$ analytically. While $\mathbb{Z}_2$-invariant observables such as $\langle M^2\rangle$ can still be estimated correctly with samples frozen in one sector, $\mathbb{Z}_2$-sensitive observables such as $\langle M\rangle$ directly test whether the two sectors are equally sampled. In Table~\ref{tab:phi4_trex_obs}, we present measurements of $\langle M\rangle$ and other sampling metrics for the methods investigated in this section. Further observables are included in App.~\ref{app:phi4_extra_obs}.
\begin{table}[t]
    \caption{Observables and sampling metrics for the $\phi^4$ target on a $48 \times 48$ lattice. The refresh time $\tau$ is the one-way trip time across the ladder divided by the number of chains, in sweeps. The magnetization $\langle M\rangle$ for HMC indicates that the chain is frozen in the negative sector, so its smaller error does not reflect this bias.}
\begin{ruledtabular}
    \begin{tabular}{lccc}
         Sampler & $\langle M\rangle$ & $\tau$ (sweeps) & Min.\ acc. \\
         \colrule
         HMC                     & -0.560(8)    & ---   & --- \\
         T-REX (uniform)         & 0.07(10) & 31(4) & 0.59 \\
         T-REX (optimal)         & 0.15(8) & 22(2) & 0.85 \\
    \end{tabular}
    \end{ruledtabular}
    \label{tab:phi4_trex_obs}
\end{table}
 
\begin{figure}
    \centering
    \includegraphics[width=0.9\linewidth]{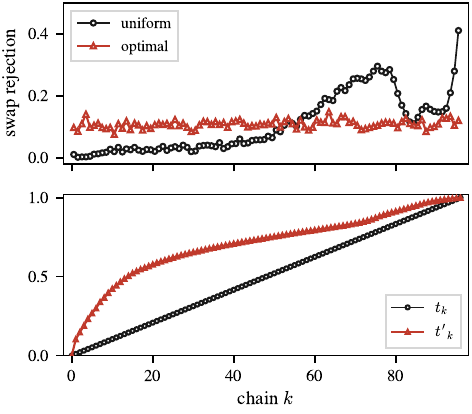}
    \caption{Comparison between the uniform (black) and optimal (red) annealing schedules for T-REX on the learned energy path in the $\phi^4$ theory on a $48 \times 48$ lattice. Top: swap rejection rate $r_k = 1-a_k$ between neighboring chains $k$, indexed from the prior. Bottom: annealing times $t_k$ of the uniform ladder and $t'_k$ of the optimal ladder for the 97 chains. The uniform ladder places chains linearly between the prior and the target, while the optimal ladder places chains more densely near the target where swaps are more difficult, which equalizes the acceptance (see top panel).}
    \label{fig:phi4_trex_ladders}
\end{figure}

Furthermore, we examine the impact of using an optimal schedule for T-REX chains. To construct the optimal ladder, we follow the two-round procedure of Ref.~\cite{Syed:2021nrpt}. In the first round, we perform T-REX on the uniform ladder described above and record the swap acceptance rate $a_k$ of each pair $(t_k, t_{k+1})$. The rejection rates $r_k = 1 - a_k$ provide an estimate of the distance between neighboring chains. A cumulative estimate up to chain $k$ is given by $D(t_k) \equiv \sum_{j < k} r_j$, which can be interpolated to provide a smooth estimate of the cumulative distance versus annealing time; for the uniform schedule, we measure a total cumulative rejection rate of 10.37(9). The new chain times $t'_k$ are then placed to achieve equal increments $D(t'_{k})-D(t'_{k-1})$, which predicts a uniform acceptance rate of $1 - (D(t'_{k})-D(t'_{k-1})) \approx 0.89$. In the second run on this ladder, with the same number of chains and cost per sweep, the swap acceptance is in fact roughly constant across annealing time, maintaining an average value of 0.891(1), as shown in Fig.~\ref{fig:phi4_trex_ladders}. The minimum measured acceptance rate is 0.85(1), compared with 0.59(2) on the uniform ladder, and the refresh time $\tau$ decreases from 31(4) to 22(2) sweeps. Configurations move along the ladder by swaps, so alleviating the bottlenecks where swaps are hardest accelerates mixing across the ladder.

\section{Conclusion and Outlook}
\label{sec:conclusion}

In this work, we present an algorithm for constructing diffusion EBMs and using them to sample lattice field theories in a systematically exact and improvable way (Algorithm~\ref{fig:main-algo}). The main premise of this work is that annealed sampling techniques can be combined with effective training strategies from diffusion models by choosing to learn the energy $E_t$ on the diffusion path.
The diffusion path is automatically equipped with a vector field that transports samples along the intermediate distributions (Eq.~\eqref{eq:diff-flow}), making flow-enhanced sampling methods (SNF and T-REX) immediately applicable. In the context of lattice field theory, these annealed sampling methods make available additional scaling directions, which mitigate the generally exponential overlap problem when sampling from an imperfect model. Sampling from a diffusion EBM using stochastic normalizing flows provides a fixed sampling efficiency when scaling the number of steps $n \sim \tilde{\Lambda}^2$, in terms of the thermodynamic path length $\tilde{\Lambda}$; sampling using T-REX gives an analogous scaling when scaling the number of replica chains $n+1 \sim \tilde{\Lambda}$ and skipping $\tau \sim \tilde{\Lambda}$ updates between saved configurations. The path length can be written in terms of relevant local metrics $\Var[\xi_t]$ for SNF sampling or $\chi_t$ for T-REX (Eq.~\eqref{eq:path-length-estimators}), allowing it to be estimated in either context.

Using analytical results and numerical applications to scalar field theory and the ${\rm O}(3)$ NLSM, we find that local-footprint models are sufficient to capture the energy $E_t$ in practical settings. These local-footprint models remain sufficient even as the lattice spacing is reduced, significantly improving the flowed path length $\tilde{\Lambda}$ relative to the unflowed $\Lambda$ and leading to one to two orders of magnitude improvement in sampling efficiency relative to annealed sampling with no flow improvement. The practical sampling complexity scales similarly to the theoretical prediction $\tilde{\Lambda} \sim \sqrt{V}$ made under assumptions of perfect independence and fixed couplings. We also identify where and how these assumptions break down---for example due to correlations between reweighting factors in SNF sampling or an anomalous scaling $\tilde{\Lambda} \sim V^{p+1/2}$ with a power $p \lesssim 0.15$ when tuning parameters along a line of constant physics.
As a final practical demonstration, we apply annealed sampling for a learned diffusion EBM in a broken-phase $\phi^4$ theory. While HMC sampling remains frozen in one mode, the learned T-REX sampler has a refresh time of approximately 30 sweeps, and tuning to the optimal annealing schedule gives a further reduction in refresh time by approximately $1.5\times$.
For other theories or diffusion schedules, it remains important to study the locality of the energy functions describing intermediate sampled distributions. Given that the diffusion path is a controlled input of the method, it could be helpful in future work to explicitly optimize for this locality such that cost-effective energy-based models can be constructed and sampled according to the annealing strategy used here.

We focus here on using a standard score-matching prescription to train our energy models to satisfy $\nabla \hat{E}_t \approx \nabla E_t$, which has the benefit of being written in terms of a simple regression loss with good training dynamics. The downside of such a loss function, however, is that the regression target has inherent noise due to stochastic estimation through either the denoising score identity in Eq.~\eqref{eq:dsi} or the target score identity in Eq.~\eqref{eq:tsi}. Combining these two loss functions mitigates this noise~\cite{debortoli:2024}, but in future work it would be interesting to investigate more sophisticated training strategies that combine regression with loss functions based on the KL divergence or the PINN loss, as has for example been proposed in Ref.~\cite{Shi:2024dps}. Such loss functions can often be constructed to converge to zero noise as training progresses. A related issue is that diffusion models require samples from the target distribution to estimate the score-matching loss function. In this work, we avoid this challenge for difficult parameters by training at smaller volumes and transferring to evaluate at larger target volumes. Because our models $\hat{E}_t$ are constructed as sums of local functions of the field degrees of freedom, this transfer is readily available. Beyond this strategy, self-learning training strategies based on imperfect sampling distributions may be of interest~\cite{Tomiya:2026tbc,Grinwald:2026dft}.

Overall, this work develops and refines a machine-learning approach which is complementary to existing flow-based, annealing, and diffusion model methods. Because the output is an annealing path with a good approximate flow transformation, these related methods are directly comparable. The methods may also be combined: an imperfect diffusion EBM can be complemented by a learned flow field which further reduces the local variance metrics, improving the sampling performance. Future work will focus on training efficacy and practicalities of sampling in more challenging lattice field theory targets, including lattice QCD. By studying a new combination of training and sampling strategies, this work takes a key step towards effective solutions to the critical slowing down problems in these contexts.

\begin{acknowledgments}
We thank Aida El-Khadra for invaluable support and mentorship throughout this project. We are grateful to Elia Cellini, Roberto Dionisio, Daniel Hackett, Saifuddin Syed, and Satria Widyanto for helpful discussions during the development of this work.
OV is supported by the U.S. Department of Energy, Office of Science under grant Contract Number DE-SC0015655 as well as by the U.S. Department of Energy, Office of Science, “High Energy Physics Computing Traineeship for Lattice Gauge Theory” Award No. DESC0024053.
GK is supported by the European Union (ERC, FoundLatt, 101212401). Views and opinions expressed are however those of the author(s) only and do not necessarily reflect those of the European Union or the European Research Council. Neither the European Union nor the granting authority can be held responsible for them.

This work has made use of the resources provided by the Edinburgh Compute and Data Facility (ECDF)~\cite{Eddie}. Some of the experiments in this research used the DeltaAI advanced computing and data resource, which is supported by the National Science Foundation (award OAC 2320345) and the State
of Illinois. DeltaAI is a joint effort of the University of Illinois Urbana-Champaign and its
National Center for Supercomputing Applications. This research used resources of the National Energy Research Scientific Computing Center (NERSC), a Department of Energy User Facility using NERSC award DDR-ERCAP 0039148. Numerical experiments and data analysis were executed using \textbf{PyTorch}~\cite{paszke:2019}, \textbf{NumPy}~\cite{harris:2020}, \textbf{SciPy}~\cite{virtanen:2019}, and \textbf{Sacred}~\cite{greff:2017}. Figures were created with \textbf{Matplotlib}~\cite{hunter:2007}.
\end{acknowledgments}

\appendix
\section{Variance-preserving diffusion}
\label{apx:euclidean_sde}
For random variables on $\mathbb{R}^n$, we consider forward diffusion SDEs modeled by the Ornstein-Uhlenbeck process
\begin{equation}\label{eq:vp_sde}
    d\phi_\lambda = -\frac{1}{2}g_\lambda^2 \phi_\lambda d\lambda + g_\lambda dW_\lambda, \quad \phi_0 \sim \rho_{\lambda=0}.
\end{equation}
The diffusivity $g_\lambda: [0, 1] \to \mathbb{R}^+$ schedules the strength of the noise added during the diffusion process.
Solving Eq.~\eqref{eq:vp_sde} first requires multiplying both sides of the SDE by an integrating factor $\mu_\lambda = \exp\left[\frac{1}{2}\int_0^\lambda g_s^2 ds\right]$ to obtain
\begin{equation}
    d(\mu_\lambda \phi_\lambda) = \mu_\lambda \Big[\big(-\cancel{\tfrac{1}{2}g_\lambda^2 \phi_\lambda} + \cancel{\tfrac{1}{2}g_\lambda^2 \phi_\lambda} \big)\, d\lambda + g_\lambda\, dW_\lambda\Big].
\end{equation}
This effects a change of variables that cancels the drift on the right hand side, leaving a pure martingale that can be integrated directly:
\begin{align}
    d\left(\mu_\lambda \phi_\lambda\right) &= \mu_\lambda g_\lambda dW_\lambda \\
    \implies \mu_\lambda \phi_\lambda - \phi_0 &= \int_0^\lambda \mu_s g_s dW_s.
\end{align}
By the It\^{o} isometry, the variance of the stochastic integral on the right-hand side is
\begin{align}
    \mathbb{E}\left[\left(\int_0^\lambda \mu_s g_s dW_s\right)^2\right] &= \int_0^\lambda \mu_s^2 g_s^2 ds \\
        &= \int_0^\lambda \frac{d}{ds}\left(\mu_s^2\right)ds \\
        &= \mu_\lambda^2 - 1,
\end{align}
so that
\begin{equation}
    \mu_\lambda \phi_\lambda - \phi_0 \sim \mathcal{N}\left(0, \mu_\lambda^2 - 1 \right).
\end{equation}
Finally, defining $\alpha_\lambda = \mu_\lambda^{-1}$, $\sigma_\lambda = \sqrt{1 - \alpha_\lambda^2}$, and using a simple reparameterization, we find that the closed form solution to Eq.~\eqref{eq:vp_sde} is
\begin{equation}\label{eq:vp_solution}
    \phi_\lambda = \alpha_\lambda \phi_0 + \sigma_\lambda\eta_\lambda, \quad \eta_\lambda \sim \mathcal{N}\left(0, \mathbb{I}\right).
\end{equation}
Notably, in addition to satisfying a ``conservation-of-variance'' law $\alpha_\lambda^2 + \sigma_\lambda^2 = 1$, the coefficients in Eq.~\eqref{eq:vp_solution} are constructed so that $\alpha_0 = 1$, $\sigma_0 = 0$, and $\alpha_1 \approx 0$, $\sigma_1\approx 1$, thereby interpolating between target configurations $\phi_0$ and samples of pure noise $\eta$. Provided that the diffusion coefficient $g_\lambda$ is sufficiently large, the density of $\phi_\lambda$ will converge exponentially closely to that of the standard normal random variable as $\lambda \to 1$.

\section{Target score identity on \texorpdfstring{$\mathbb{S}^2$}{S2}}
\label{app:o3-tsi}
In this section, we generalize the Lie group target score identity provided in Ref.~\cite{debortoli:2024} to the case of the unit sphere $\mathbb{S}^2$. Here, we denote elements of the 2-sphere as $\vec{x},\vec{y} \in \mathbb{S}^2$ to abstractly represent unit vectors embedded in $\mathbb{R}^3$, such as the spin configurations of the ${\rm O}(3)$ NLSM.

We first introduce the exact control variate
\begin{equation}
\begin{aligned}
    C(\vec{y}, \vec{x}) &= R_{\vec{x} \to \vec{y}}[\nabla_{\vec{x}} \log K_\lambda(\vec{y}|\vec{x}) - \nabla_{\vec{x}} S(\vec{x})] \\
    &\quad + \mathrm{div}_{\vec{x}} R_{\vec{x} \to \vec{y}}.
\end{aligned}
\end{equation}
In terms of the conditional distribution
\begin{equation}
    p(\vec{x} | \vec{y}) \propto e^{-S(\vec{x})} K_\lambda(\vec{y} | \vec{x}),
\end{equation}
this control variate is a total derivative,
\begin{equation}
    \mathbb{E}_{\vec{x} | \vec{y}}[C(\vec{y}, \vec{x})] = \int d\mu(\vec{x}) \, \mathrm{div}_{\vec{x}} \left[ p(\vec{x}|\vec{y}) R_{\vec{x}\to\vec{y}} \right] = 0.
\end{equation}
The tangent-space map $R_{\vec{x} \to \vec{y}}$ sends the tangent space at $\vec{x}$ to the tangent space at $\vec{y}$. A useful choice is the divergence-free map
\begin{equation}
    R_{\vec{x} \to \vec{y}} [\vec{v}] = \vec{y} \times (\vec{v} \times \vec{x}).
\end{equation}
Under this choice, the total target score identity becomes
\begin{equation} \label{eq:o3-tsi}
\begin{aligned}
    \mathcal{T}_\lambda(\vec{y}, \vec{x}) &= \nabla_{\vec{y}} \log K_\lambda(\vec{y}|\vec{x}) + C(\vec{y},\vec{x}) \\
    &= \cancel{\nabla_{\vec{y}} \log K_\lambda(\vec{y}|\vec{x})} \\
    &\quad + R_{\vec{x} \to \vec{y}}[\cancel{\nabla_{\vec{x}} \log K_\lambda(\vec{y}|\vec{x})} - \nabla_{\vec{x}} S(\vec{x})] \\
    &= - \vec{y} \times (\nabla_{\vec{x}} S(\vec{x}) \times \vec{x}).
\end{aligned}
\end{equation}
The cancellation in the second-to-last equation comes from the rotational invariance of the $\mathbb{S}^2$ heat kernel and symmetry with respect to its two arguments, which leads to
\begin{equation}
    R_{\vec{x} \to \vec{y}}[\nabla_{\vec{x}} \log K_\lambda(\vec{y}|\vec{x})] = -\nabla_{\vec{y}}\log K_\lambda(\vec{y}|\vec{x}).
\end{equation}

For applications to the O(3) lattice field theory, the same tangent space identity can be extended to O(3) fields $\vec{\phi}_\lambda$ and $\vec{\phi}_0$ by taking the action $S(\cdot)$ to act on the full configuration and evaluating Eq.~\eqref{eq:o3-tsi} independently over all sites.

\section{Further metrics}\label{app:scaling}

\begin{figure}
    \centering
    \includegraphics[width=0.9\linewidth]{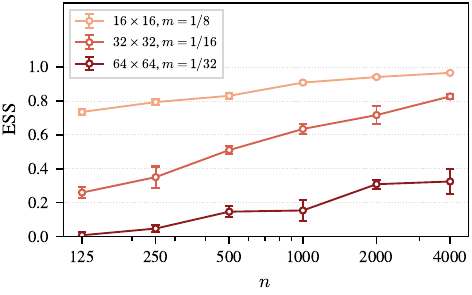}
    \caption{Scaling of the effective sample size with the number of annealing steps $n$ for the free scalar field theory.}
    \label{fig:ess_free}
\end{figure}

In Fig.~\ref{fig:ess_free}, we plot representative examples of the scaling behavior of the ESS with the number of SNF annealing steps $n$. For every $L$, we observe that the ESS rises monotonically with the number of annealing steps, indicating that our approach enables systematic refinement of sample quality post-training. In approaching the quasi-static limit, the variance of the log-reweighting factors monotonically shrinks as $n \to \infty$.

\begin{figure}
    \centering
    \includegraphics[width=0.9\linewidth]{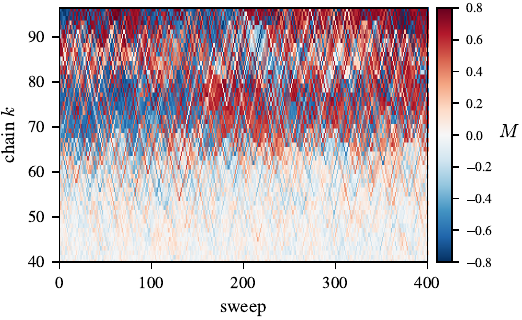}
    \caption{Magnetization $M$ for the $48 \times 48$ $\phi^4$ configurations on chains $k=40$--96 of the $97$-chain optimal ladder. The first 400 of 2000 T-REX sweeps are shown.}
    \label{fig:phi4_sign_map}
\end{figure}

To further explore the better mixing behavior in the $\phi^4$ application presented in Sec.~\ref{sec:phi4_trex_demo}, we measure the average magnetization $M$ across all chains during T-REX evaluation. The result is plotted in Fig.~\ref{fig:phi4_sign_map}, in which we observe significantly higher magnetization of samples near the target chain and almost no magnetization over roughly the first half of the chains. This matches the expectation that the learned annealing path defined by the diffusion schedule crosses from the broken phase to the symmetric phase as more noise is added (i.e., moving from the top chain towards the bottom in the figure). We also observe an apparent defect of the learned model, given by the region of decreased magnetization roughly in chains 80--90, which indicates that sampling efficiency could be further improved by more complex model parameterizations or more training. Regardless, the plot explains the ability of this T-REX sampler to overcome the freezing problem effectively: individual samples propagate rapidly from the noise chain in the symmetric phase to the target chain in the broken phase.

\section{Observables}\label{app:observables}

\subsection{Free scalar observables}
For the free scalar field theory, we measure the average value of the squared field $\langle \phi^2 \rangle$. The analytical value of this quantity is
\begin{equation}
    \langle \phi^2\rangle = \frac{1}{V}\sum_k \frac{1}{\omega(k)^2},
\end{equation}
which we compare against measurements from SNF, REX, and T-REX in Table~\ref{tab:free_obs}.
\begin{table}
\caption{Measurements of $\langle\phi^2\rangle$ in the free scalar theory ($m = 2/L$) from SNF (reweighted, $n=4000$), REX, and T-REX, compared with exact values for each $L$. Errors are bootstrap estimates (blocks of 100 sweeps for REX/T-REX).}
\begin{ruledtabular}
\begin{tabular}{lcccc}
    $L$ & Exact & SNF & REX & T-REX \\
    \hline
    16 & 0.7247 & 0.742(27) & 0.740(12) & 0.720(8) \\
    32 & 0.8357 & 0.789(18) & 0.822(15) & 0.817(13) \\
    64 & 0.9462 & 0.969(67) & 1.091(74) & 0.958(46) \\
\end{tabular}
\end{ruledtabular}
\label{tab:free_obs}
\end{table}
 Results are consistent with the exact values, with all deviations at most $2.6\sigma$.

\subsection{\texorpdfstring{$\phi^4$}{phi4} observables}\label{app:phi4_extra_obs}
In Sec.~\ref{sec:phi4_trex_demo}, we applied T-REX sampling to the $\phi^4$ theory and measured the average magnetization $\langle M\rangle$ in Table~\ref{tab:phi4_trex_obs} to demonstrate improved mixing. In this section, we include additional $\phi^4$ observables as a further check of our method, comparing measurements to those obtained on a reference HMC ensemble and a baseline REX calculation. Specifically, we compute three observables: the absolute magnetization
\begin{equation}
    \langle|M|\rangle = \frac{1}{V} \left \langle \left| \sum_x \phi(x) \right| \right\rangle,
\end{equation}
the magnetic susceptibility
\begin{equation}
    \chi = \langle M^2\rangle,
\end{equation}
and the Binder cumulant,
\begin{equation}
    U_4 = 1 - \frac{\langle M^4\rangle}{3 \langle M^2\rangle^2}.
\end{equation}

\begin{table}
\caption{Observables for the $\phi^4$ theory on a $48\times48$ lattice from REX and T-REX with the symmetrized learned energies, compared with a standard HMC simulation. Uncertainties are obtained by bootstrap resampling over the 97 independent chains (HMC) or resampling over blocks of 100 sweeps (REX, T-REX).}
\begin{ruledtabular}
\begin{tabular}{lccc}
    Sampler & $\langle |M|\rangle$ & $\chi$ & $U_4$ \\
    \hline
    HMC & 0.5537(28) & 0.3133(22) & 0.6468(19) \\
    REX (uniform) & 0.5513(84) & 0.3113(79) & 0.6444(42) \\
    T-REX (uniform) & 0.5527(52) & 0.3130(49) & 0.6458(24) \\
    T-REX (optimal) & 0.5471(47) & 0.3091(45) & 0.6410(33) \\
\end{tabular}
\end{ruledtabular}
\label{tab:phi4_obs}
\end{table}

\begin{table}
\begin{ruledtabular}
\begin{tabular}{llcc}
$(\beta, L)$ & Method & $\langle \vec\phi(x)\cdot\vec\phi(x+\hat\mu)\rangle$ & $\xi$ \\
\hline
\multirow{3}{*}{$(1.4, 16)$} & cluster        & 0.56815(19) & 5.825(14) \\
                             & SNF ($n=1000$) & 0.5693(17)  & 5.82(16) \\
                             & T-REX ($n=50$) & 0.5677(14)  & 5.62(11) \\
\hline
\multirow{3}{*}{$(1.5, 32)$} & cluster        & 0.60318(9)  & 10.120(25) \\
                             & SNF ($n=1000$) & 0.6017(9)  & 10.58(32) \\
                             & T-REX ($n=50$) & 0.6025(7)   & 9.70(33) \\
\hline
\multirow{3}{*}{$(1.6, 64)$} & cluster            & 0.63602(4)  & 17.99(5) \\
                             & SNF ($n=500$)      & 0.6363(7)   & 16.2(1.1) \\
                             & T-REX ($n=50$)     & 0.6362(8)   & 16.9(7) \\
\end{tabular}
\end{ruledtabular} \caption{Observables for the ${\rm O}(3)$ model compared with a cluster-algorithm reference ($10^5$ configurations). Shown are the nearest-neighbor spin correlation and the second-moment correlation length $\xi$. Errors are bootstrap estimates. At (1.6, 64), the T-REX averages use only the last 200 MCMC steps to avoid the slow thermalization of the correlation function.}
\label{tab:o3_obs}
\end{table}

\begin{figure}
    \centering
    \includegraphics[width=0.9\linewidth]{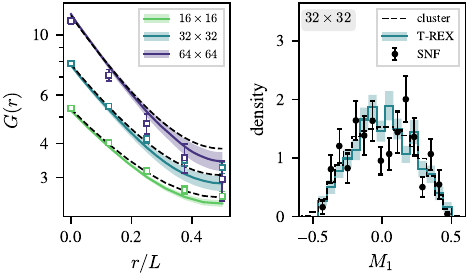}
    \caption{Comparison of ${\rm O}(3)$ observables with the cluster-algorithm reference (black dashed lines). Left: projected two-point correlation function $G(r)$. T-REX results ($n=50$) are shown at all separations as a line with a $\pm 1\sigma$ band; reweighted SNF results ($n=1000$ for $L=16,32$ and $n=500$ for $L=64$) are shown as points at every $L/8$ separation. Right: distribution of the average first spin component $M_1$ at $(1.5, 32)$ from T-REX with $n=100$ and SNF with $n=1000$. Errors are block-bootstrap estimates.}
    \label{fig:o3_observables}
\end{figure}

Measurements across HMC, REX, and T-REX (uniform and optimal schedules) are collected in Table~\ref{tab:phi4_obs}. All replica exchange results were obtained using the local-footprint energy model trained at $16 \times 16$, transferred to $48 \times 48$, and symmetrized according to Eq.~\eqref{eq:symmetrize_model}. 

We observe good agreement across all models for the three observables. The Binder cumulant $U_4$ is close to its expected value ($2/3$) for two sharp peaks in $M$, confirming that the $\phi^4$ target is deep in the broken phase. Notably, HMC fails to tunnel and REX never transports configurations across the full ladder, but the observables still agree with the T-REX measurements. This is because all three observables are $\mathbb{Z}_2$-invariant, and can therefore be estimated correctly even if configurations are frozen in one of the two sectors in the broken phase.

\subsection{O(3) observables}
For the ${\rm O}(3)$ NLSM, we compare measurements of four observables obtained using the methods we presented in this work against those from a reference code using a cluster algorithm~\cite{Swendsen:1987ce,Wolff:1988uh}. The first is the nearest-neighbor spin correlation $\langle \vec{\phi}(x) \cdot \vec{\phi}(x + \hat{\mu})\rangle$, averaged over lattice sites $x$ and spacetime directions $\mu$. We also compute the zero-momentum projected two-point correlation function 
\begin{equation}
    G(r) = \frac{1}{V} \sum_{\substack{x,y \\ y_1 = x_1 + r}} \left\langle \vec{\phi}(x) \cdot \vec{\phi}(y) \right\rangle.
\end{equation}
From its Fourier transform $\tilde{G}(k) = \sum_r G(r) e^{i kr}$, the second-moment correlation length is defined as
\begin{equation}
    \xi = \frac{1}{2\sin(\pi/L)}\sqrt{\frac{\tilde{G}(0)}{\tilde{G}(2\pi/L)} - 1}.
\end{equation}
Finally, we compare the distributions of the lattice-averaged first spin component,
\begin{equation}
    M_1 = \frac{1}{V} \sum_x \phi^1(x),
\end{equation}
which is symmetric about zero in the target theory.

Results for the first two observables are collected in Table~\ref{tab:o3_obs}. SNF estimates are reweighted using the importance weights of the generated samples, while T-REX estimates are averages over the target chain. Both the nearest-neighbor correlation and the correlation length agree with the cluster reference within $2\sigma$ for every method and lattice size.

The full correlation functions and the $M_1$ distribution are shown in Fig.~\ref{fig:o3_observables}. T-REX reproduces the cluster-algorithm $G(r)$ within errors at all separations on both the $32 \times 32$ and $64 \times 64$ lattices, and the $M_1$ distributions from both methods agree with the reference at $32 \times 32$.

\end{document}